\documentclass[twocolumn]{aastex62}

\graphicspath{{./}{figures/}}

\received{July, 2020}
\revised{}
\accepted{}
\submitjournal{ApJL}

\shorttitle{Intermediate-mass stars become magnetic white dwarfs}
\shortauthors{Caiazzo et al.}
\begin{document}

\title{Intermediate-Mass Stars Become Magnetic White Dwarfs}

\correspondingauthor{Ilaria Caiazzo}
\email{ilariac@caltech.edu}

\author[0000-0002-4770-5388]{Ilaria Caiazzo}
\altaffiliation{Sherman Fairchild Fellow}
\affiliation{Walter Burke Institute for Theoretical Physics, \\
TAPIR, Caltech, \\ 
Mail Code 350-17, Pasadena, CA 91125, USA}
\affiliation{University of British Columbia,\\
Department of Physics and Astronomy,\\
Vancouver, BC V6T 1Z1, Canada}

\author[0000-0001-9739-367X]{Jeremy Heyl}
\affiliation{University of British Columbia,\\
Department of Physics and Astronomy,\\
Vancouver, BC V6T 1Z1, Canada}

%% The \author command can take an optional ORCID.
\author[0000-0001-9002-8178]{Harvey Richer}
\affiliation{University of British Columbia,\\
Department of Physics and Astronomy,\\
Vancouver, BC V6T 1Z1, Canada}

\author[0000-0001-7453-9947]{Jeffrey Cummings}
\affiliation{Space Telescope Science Institute,\\
3700 San Martin Drive, Baltimore, MD 21218, USA}
\affiliation{Johns Hopkins University,\\
Department of Physics and Astronomy,\\
3400 N. Charles Street, Baltimore, Maryland 21218, USA}

\author[0000-0002-7022-8380]{Leesa Fleury}
\affiliation{University of British Columbia,\\
Department of Physics and Astronomy,\\
Vancouver, BC V6T 1Z1, Canada}

\author{James Hegarty}
\affiliation{University of British Columbia,\\
Department of Physics and Astronomy,\\
Vancouver, BC V6T 1Z1, Canada}

\author[0000-0001-9690-4159]{Jason Kalirai}
\affiliation{Applied Physics Laboratories,\\
Johns Hopkins University,\\
Baltimore, MD 21218, USA}

\author[0000-0002-6549-9792]{Ronan Kerr}
\affiliation{University of Texas at Austin,\\
Department of Astronomy,\\
Austin, Texas 78712-1205, USA}

\author[0000-0001-7442-6926]{Sarah Thiele}
\affiliation{University of British Columbia,\\
Department of Physics and Astronomy,\\
Vancouver, BC V6T 1Z1, Canada}

\author[0000-0001-9873-0121]{Pier-Emmanuel Tremblay}
\affiliation{University of Warwick,\\
Department of Physics,\\
Coventry, CV4 7AL, UK}

\author{Michael Villanueva}
\affiliation{University of British Columbia,\\
Department of Physics and Astronomy,\\
Vancouver, BC V6T 1Z1, Canada}



\begin{abstract}

When a star exhausts its nuclear fuel, it either explodes as a supernova or more quiescently becomes a white dwarf, an object about half the mass of our Sun with a radius of about that of the Earth. About one fifth of white dwarfs exhibit the presence of magnetic fields, whose origin has long been debated as either the product of previous stages of evolution or of binary interactions. We here report the discovery of two massive and magnetic white dwarf members of young star clusters in the Gaia DR2 database, while a third massive and magnetic cluster white dwarf was already reported in a previous paper. These stars are most likely the product of single-star evolution and therefore challenge the merger scenario as the only way to produce magnetic white dwarfs. The progenitor masses of these stars are all above 5 solar masses, and there are only two other cluster white dwarfs whose distances have been unambiguously measured with Gaia and whose progenitors' masses fall in this range. This high incidence of magnetic white dwarfs indicates that intermediate-mass progenitors are more likely to produce magnetic remnants and that a fraction of magnetic white dwarfs forms from intermediate-mass stars. 

%As there are only five such white dwarfs known, including the three newly discovered ones, such a high incidence of magnetic objects indicates that most intermediate-mass progenitors produce magnetic remnants.

\end{abstract}



\section{Introduction} \label{sec:intro}
White dwarfs (WDs) are all born in the same manner, as the compact remnants of low and intermediate-mass stars up to about eight times the mass of our Sun. However, some WDs are peculiar, as their surfaces are threaded with magnetic fields that range from a few thousand to a billion Gauss. A recent study \citep{2019A&A...628A...1L} that focused on DA (hydrogen atmosphere) WDs showed that about $20\pm5$\% of the 80 DA WDs that are within 20~pc of the Sun are magnetic, of which only seven ($\sim9$\%) have magnetic fields exceeding 1 MG. If we extend the sample to all the WDs within 20~pc of the Sun \citep{2018MNRAS.480.3942H}, we find a similar result: of the 126 WDs that have a spectral classification, about 20\% (23) are magnetic and about 10\% (13) have magnetic fields above 0.5~MG. The origin of magnetism in WDs is still a matter of debate, and the theories that have been suggested fall into two main categories. In one model, the binary scenario, magnetic fields in WDs are created by the convective dynamos that arise during a common envelope phase in the interaction with a companion star  \citep{2008MNRAS.387..897T,2011PNAS..108.3135N,2012ApJ...749...25G,2014MNRAS.437..675W}. In this case, a magnetic WD would either be the product of a merger or would still be in a close binary. In the main alternative hypothesis, the magnetic fields found in WDs are thought to be of fossil origin, that is, they were already present in the progenitor star or somehow were generated in the different evolutionary stages that precede the formation of the WD. In this second scenario, the progenitors could be stars that already displayed strong magnetic fields, such as the magnetic main-sequence Ap/Bp stars  \citep{1964ApJ...140.1309W,2003A&A...403..693M,2004MNRAS.355L..13T}, or the magnetic field could be hidden in the progenitor stars' core, as is suggested by recent studies on asteroseismology of red giant stars
 \citep{2015Sci...350..423F,2016Natur.529..364S,2016ApJ...824...14C}.

It is well known that magnetic WDs are on average more massive than non-magnetic ones \citep{2020arXiv200110147F,2020arXiv200110672K,2020arXiv200600874M}, and this could either indicate that the mass of the progenitor star is somehow related to the genesis of the magnetic field, as more massive stars create more massive WDs, or that magnetic WDs are the products of WD mergers, or both. We here report the discovery of two DA WDs from the Gaia DR2 catalog \citep{2016A&A...595A...1G,2018A&A...616A...1G} in young open star clusters that exhibit the presence of a magnetic field on their surface. The WDs are located in Messier 39 (M 39, NGC 7092) and ASCC 47. We already reported the discovery of a third magnetic DB (helium atmosphere) WD in Messier 47 (M 47) in \citet{2019ApJ...880...75R}, hereafter Paper I. ASCC 47 is the youngest open cluster ($90\pm20$ Myr) known to contain a WD, and the WD itself is the youngest and hottest found in such a cluster; M~39 is $280\pm20$ Myr old and M 47 is $150\pm20$ Myr old (see below).  As such young clusters could have only produced massive WDs above $\sim0.9$~M$_\odot$  \citep{2018ApJ...866...21C}, a merger inside the cluster would have created a remnant that is above the maximum mass for a WD \citep[the Chandrasekhar limit of $1.38$~M$_\odot$,][]{1987ApJ...322..206N}, and therefore would have exploded in a type Ia supernova \citep{2015ApJ...805L...6S}. Furthermore, the fraction of binaries among the most massive stars in the cluster is low.  For these reasons, the newly discovered WDs rule out the double-degenerate merger scenario as the only channel for the formation of magnetic white dwarfs.

The fact that the three WDs are members of star clusters means that the mass of their progenitor stars can be inferred from the age of the cluster and from the cooling time of the WD. We find that the progenitor stars of the three WDs all had masses above 5~M$_\odot$, and there are only two other cluster WDs known in the Gaia-DR2 database whose progenitors were this massive. This high incidence of magnetic WDs from intermediate-mass progenitors suggests that more massive stars are more likely to produce magnetic WDs, which would also explain why magnetic WDs are on average more massive.

\section{Identification of the objects}

The publication of more than a billion high-precision parallaxes and proper motions in Gaia DR2 \citep{2016A&A...595A...1G,2018A&A...616A...1G} has been revolutionary for many branches of stellar astrophysics. In the case of nearby open clusters, which can extend up to several square degrees on the sky, the identification of cluster members via the measurement of parallaxes and proper motions over large parts of the sky was not feasible previously for many clusters but has become straightforward with Gaia. The three WDs analyzed in this paper were discovered as part of a larger survey of young open clusters whose purpose is to find massive cluster white dwarfs \citep{Richerteal2020}. The WDs are identified as cluster members with high confidence because they lie inside cluster boundaries in coordinate space and have parallaxes and proper motions similar to those of their respective clusters, as can be seen in Fig.~\ref{fig:cluster_and_WD} (for more details on each object see Appendix~\ref{sec:membership} and Paper I). 

To confirm the white-dwarf nature of the candidates, and to provide an estimate of their temperatures and surface gravities, we obtained their spectra with GMOS in long-slit mode on Gemini North (for the WD in M 39) and Gemini South (for the WD in ASCC 47), exploiting its fast turnaround program. The spectrum for the WD in M 47 was already obtained and published in Paper I. We used the B600 grating centered at 512 and 508 nm (for dithering) with no blocking filter.  The slit width, set to 1'', provided about 5~\AA\ resolution. We used IRAF \citep{1986SPIE..627..733T} for reduction.
\begin{figure*}
    \centering
    \includegraphics[width=\textwidth]{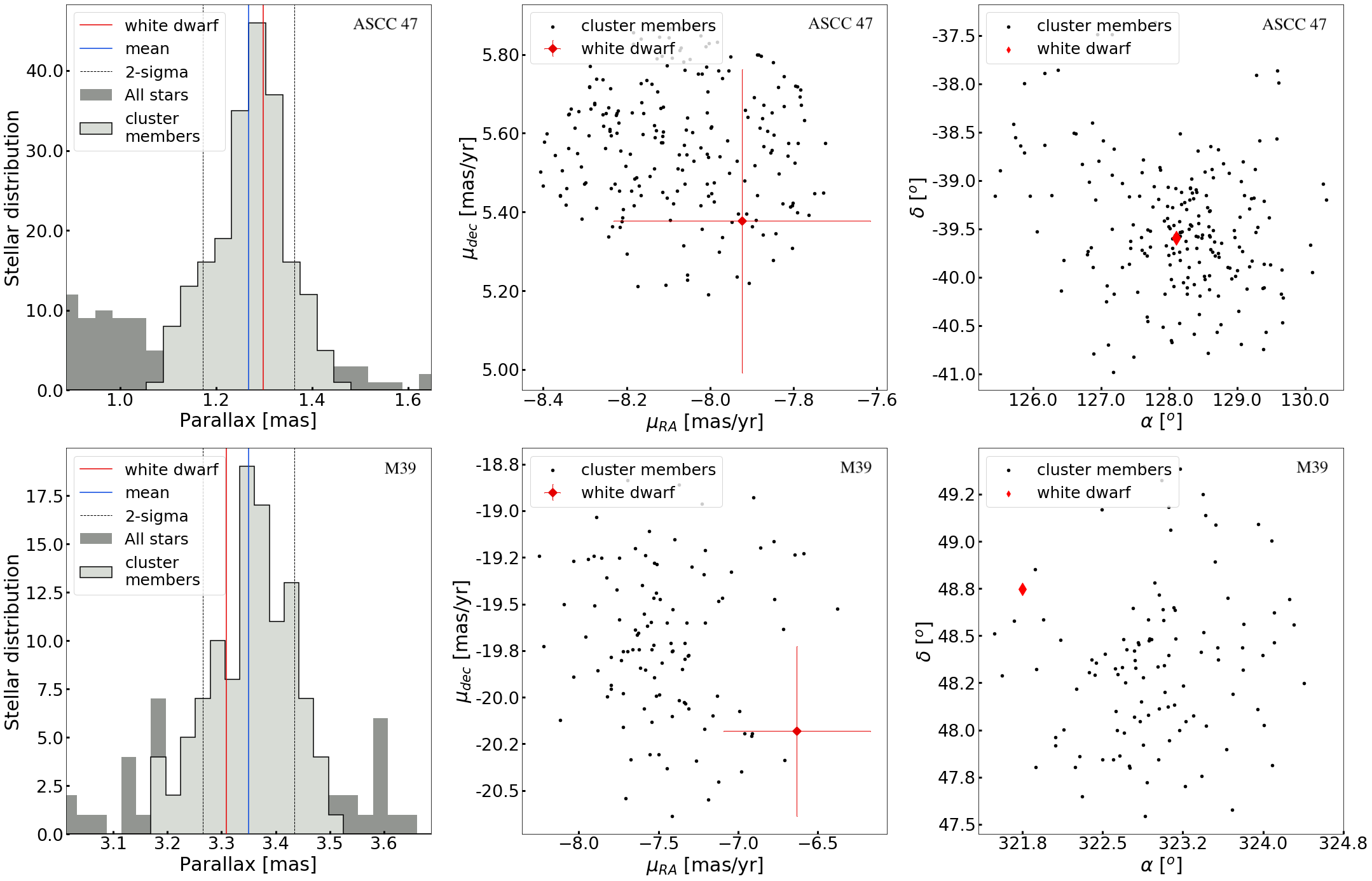}
    \caption{The phase space cuts to determine the cluster membership: parallax, proper motion and position on the sky. In the left panels, the grey histogram shows the distribution of parallaxes centered on the directions of ASCC 47 and M 39 within 2.5 and 1.5 degrees of the centers of the two clusters. The blue line indicates the mean cluster values (1.273 and 3.350 mas), while the dashed lines delineate the $\pm2\sigma$ parallax limits. The solid red line provides the measured parallax of the WD. The middle panels show the proper motions of the stars with the WD in red.  The right panel shows the sky positions of the stars and the WD (again in red).}
    \label{fig:cluster_and_WD}
\end{figure*}

\section{Properties of the white dwarfs}
\subsection{Magnetic field}
\begin{figure*}[tb]
    \centering
    \includegraphics[width=0.48\textwidth]{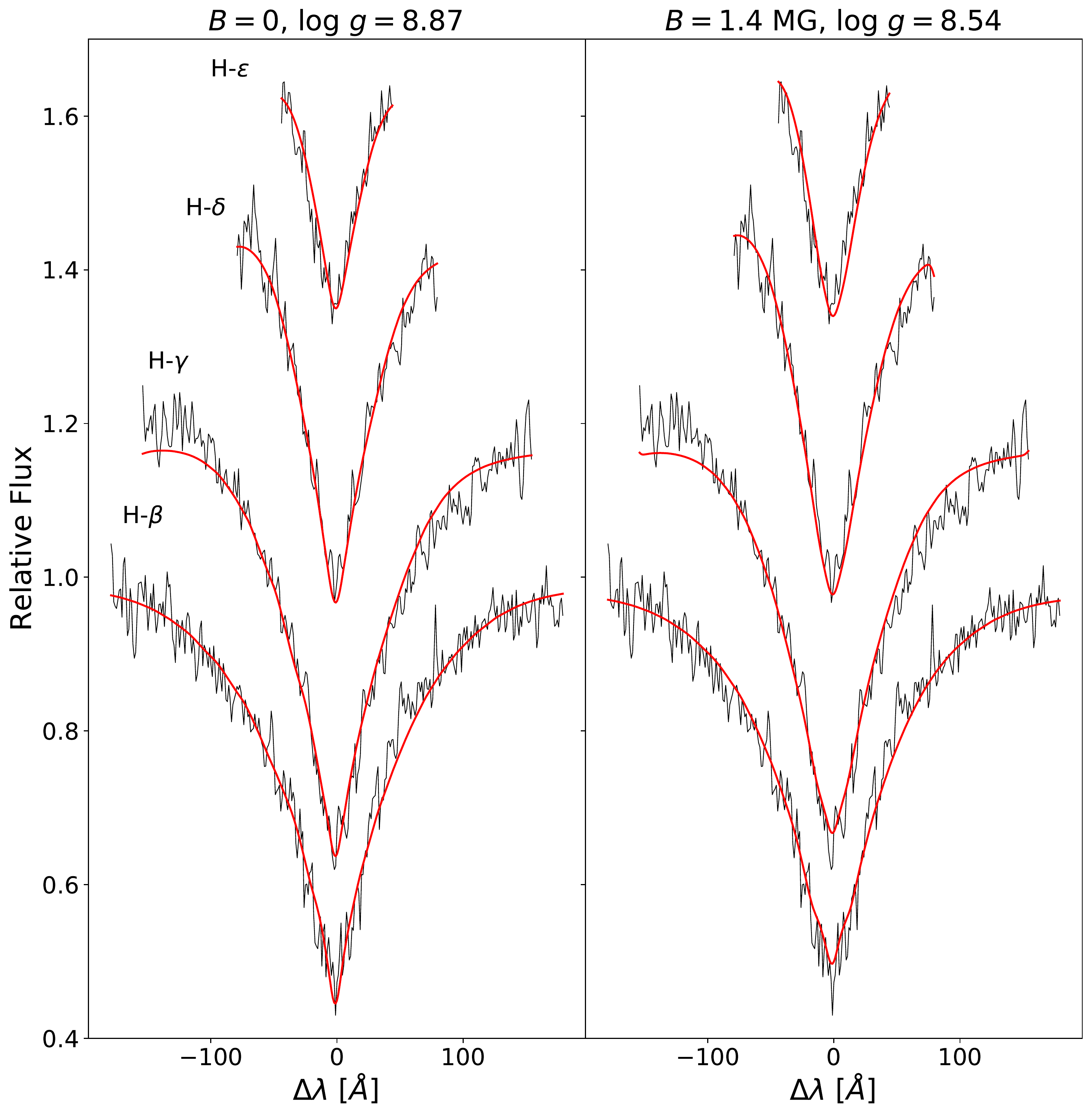}
    \includegraphics[width=0.48\textwidth]{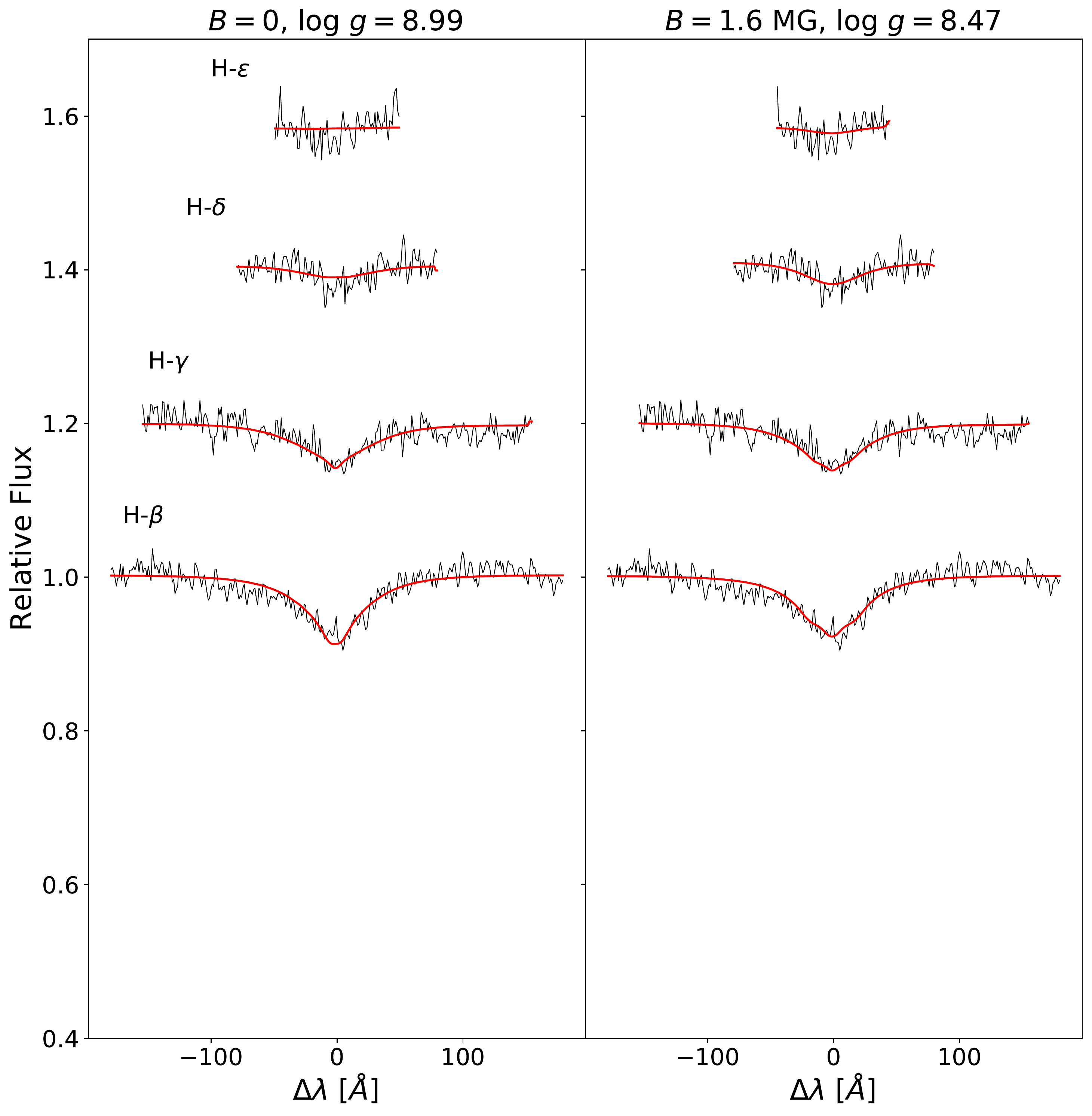}
    \includegraphics[width=0.48\textwidth]{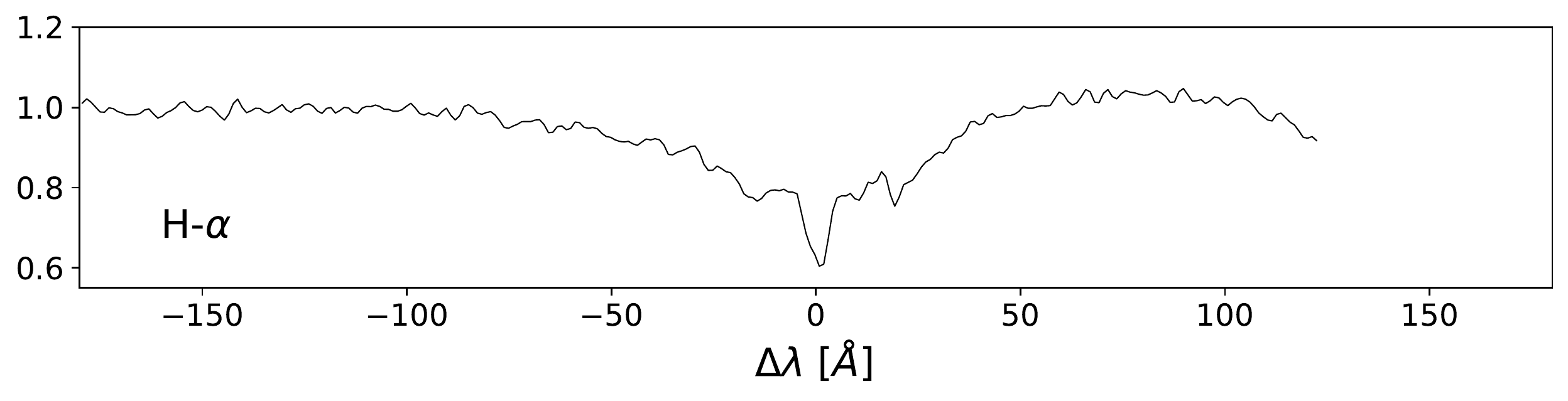}
    \includegraphics[width=0.48\textwidth]{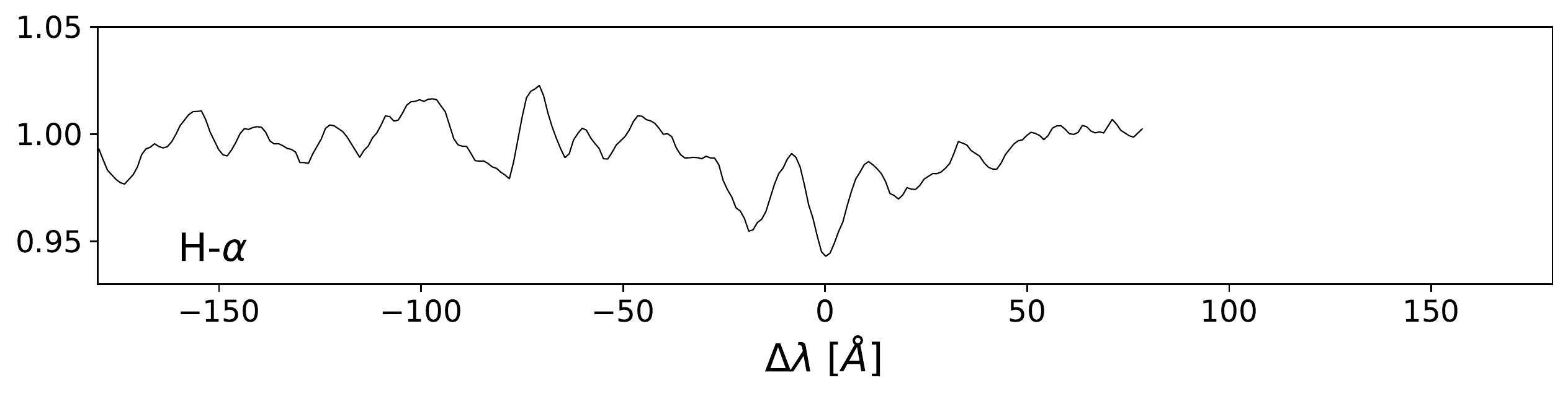}
    \caption{Upper: Balmer lines (H$\beta$ to H$\epsilon$) in the normalised spectra of M 39 (left) and ASCC 47 (right). Higher-order lines than H$\beta$ have been shifted up by 0.2 each. In each panel, in red, the left side shows the best fit with the non-magnetic models and the right side with magnetic models. The values for the best fits are given in the text. Lower: H$\alpha$ in M 39 (left) and ASCC 47 (right). } 
    \label{fig:lines}
\end{figure*}

The spectra for the objects in ASCC 47 and M 39 are consistent with hot, massive DA WDs: the only spectral features are the hydrogen Balmer lines, which are very broad. In Fig.~\ref{fig:lines} the lines have been plotted using equal ranges on the x and y axis, so it is possible to compare the two WDs; the lines for the ASCC 47 WD are extremely weak because the object is very hot. In the lower panels of Fig.~\ref{fig:lines} we can see the H$\alpha$ lines for both objects, which show Zeeman splitting. For magnetic fields lower than about 10 MG, the effect of the magnetic field reduces to the linear Zeeman effect, for which the Balmer lines split into three components, with a central component at the same wavelength as the zero-field line and with a blueshifted and a redshifted component. 
%If the lines are intrinsically narrow and the magnetic field strong, the Zeeman splitting is evident because all the lines appear as triplets, as can be clearly seen in the spectrum of the magnetized WD in M 47 (see Paper I). When the surface gravity is high and the WD hot, and the lines are intrinsically broad and weak, the splitting is only visible in H$\alpha$ and is not obvious for higher-order lines, as the triplet blends together into an even broader line. 
The separation between the H$\alpha$ Zeeman components in both objects is about 20~\AA, which corresponds to a field of about 1~MG. For the WD in M 47, the Zeeman splitting of the helium absorption lines indicates a magnetic field strength of 2.5~MG (see Paper I). As the observed Zeeman splitting represents the mean field across the surface of the star, the value of the polar magnetic field is bound to be higher.

\subsection{Temperature, surface gravity and mass}
\begin{figure}[tb]
    \centering
    \includegraphics[width=\columnwidth]{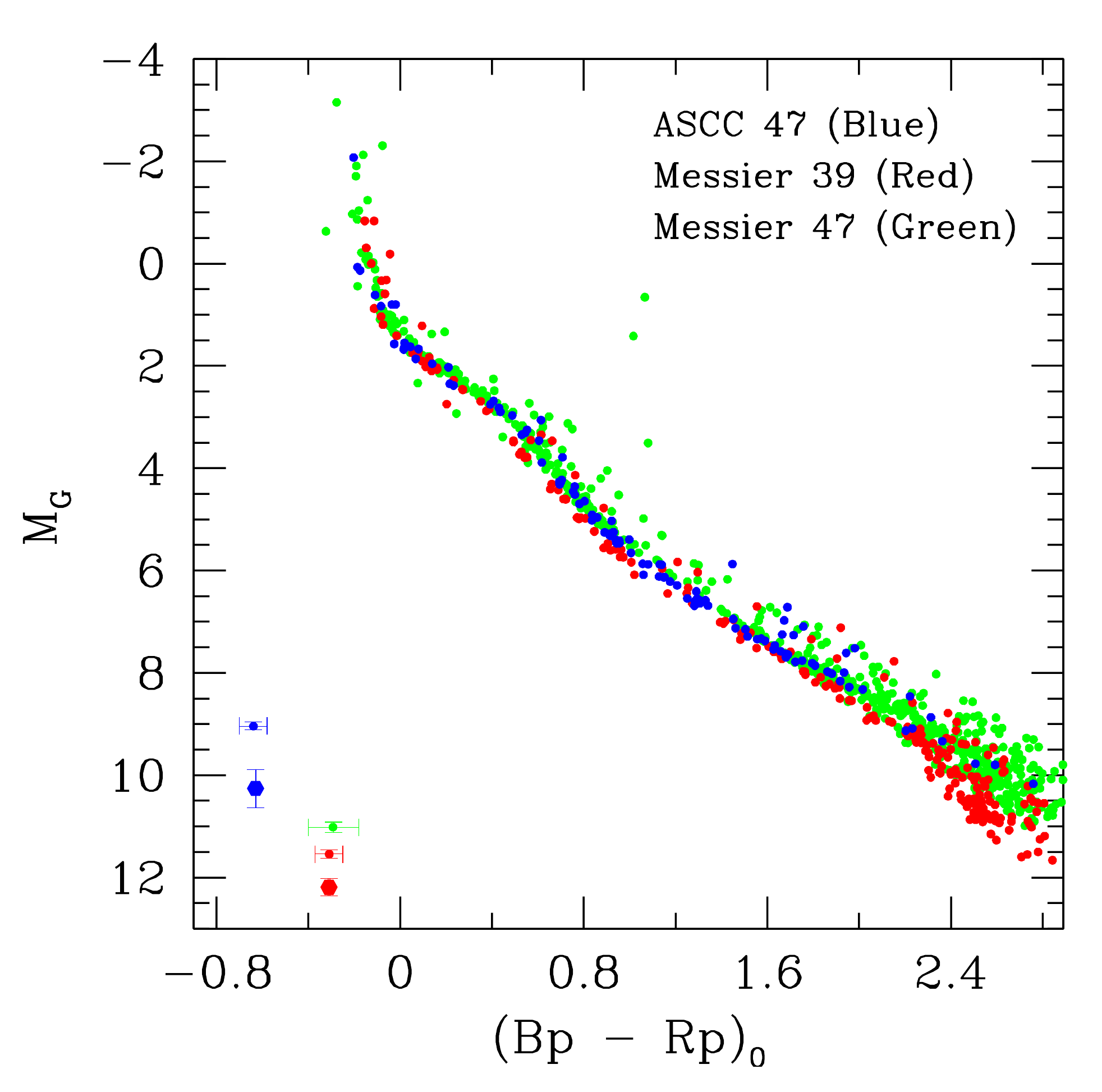}
    \caption{Gaia color-magnitude diagram for the three clusters. The WDs can be seen in the lower left corner. The observed values are indicated as small dots with error bars (from Gaia). For M 39 and ASCC 47, the hexagonal shape indicates the expected location in the CMD of the WDs for the $\log g$ and $T_{\rm{eff}}$ found from spectroscopy using the models by Tremblay et al. \citep{2011ApJ...730..128T} (the error bars come from 1-$\sigma$ errors in the fit).} 
    \label{fig:cmd}
\end{figure}

The left panels of Fig~\ref{fig:lines} show the fit of non-magnetic atmosphere models to four Balmer lines: H$\beta$, H$\gamma$, H$\delta$ and H$\epsilon$. The fitting routine is explained in Appendix~\ref{sec:fitting} and returns the following best fitting values for the logarithm of the surface gravity ($\log g$) and for the effective temperature ($T_{\rm{eff}}$): $\log g =8.87\pm0.07$~$[\log(\rm{cm}~\rm{s}^{-2})]$ and $T_{\rm{eff}}=18,400\pm300$~K for the WD in M 39, and $\log g =8.99\pm0.13$~$[\log(\rm{cm}~\rm{s}^{-2})]$ and $T_{\rm{eff}}=116,000\pm3,000$~K for ASCC 47. For both stars, the non-magnetic fit returns a surface gravity that is too high if we take into account the brightness of the WDs. In Fig.~\ref{fig:cmd}, we plot the observed absolute magnitudes for the WDs (small points with error bars) and the magnitudes expected for WDs with the temperatures and surface gravities obtained from spectroscopy (hexagons with error bars). The discrepancy lies in the expected radii for the given surface gravities: comparing synthetic photometric models \citep{2001ApJS..133..413B,2006AJ....132.1221H,2006ApJ...651L.137K,2011ApJ...730..128T,2018ApJ...863..184B} to the photometry available in Gaia, VPHAS+ \citep{2014MNRAS.440.2036D} and Pan-STARRS \citep{2016arXiv161205560C}, we measure the WD radii to be $6,100\pm150$~km for M 39 and $6,750\pm250$~km for ASCC 47. Using the $\log g$ that we found from spectroscopy, these radii would imply an unreasonable mass of $2.1$~M$_\odot$ for M 39 and $3.4$~M$_\odot$ for ASCC 47, well above the Chandrasekhar limit of $1.38$~M$_\odot$ \citep{1987ApJ...322..206N}. This discrepancy is due to the magnetic field: the effect of surface gravity and magnetic field are degenerate in broadening the lines.

%Another way to look at this discrepancy is that the surface gravity measured spectroscopically would imply, when a mass-radius relation for a typical WD is employed, a radius of about 4000~km, much smaller than indicated by the photometry. The Balmer lines thus must be broadened by something in addition to gravity alone. The presence of a magnetic field on the surface of the WDs, which is also suggested by the fact that the line cores in both objects are broad (gravity broadens mainly the wings), can explain the excess of broadening. The presence of magnetism is also hinted by the peculiar shape of the H$\alpha$ lines in both WDs, shown in the lower panels of Fig.~\ref{fig:lines}. Our signal-to-noise in this range of wavelengths is too low to say anything definitive (especially for ASCC 47), however the triplet shape of H$\alpha$ in both WDs seems to indicate the presence of Zeeman splitting. 

Because of this degeneracy, we cannot measure the surface gravity from spectroscopy and we therefore use photometric fitting \citep[as in][see Appendix~\ref{sec:phot}]{2019MNRAS.482.4570G} . We obtain the following values:
$\log g =8.54\pm0.04$~$[\log(\rm{cm}~\rm{s}^{-2})]$ for M 39, and $\log g =8.47\pm0.05$~$[\log(\rm{cm}~\rm{s}^{-2})]$ for ASCC 47, which yield a mass of $0.95\pm0.02$~M$_\odot$ for M 39 and $1.01\pm 0.02$~M$_\odot$ for ASCC 47. The photometric fit for the WD in M 47 returns a mass of $1.06\pm0.05$~M$_\odot$. 

The right panels of Fig.~\ref{fig:lines} show fits with simple magnetized atmospheric models that include the linear Zeeman effect assuming the field to be constant over the surface (see Appendix~\ref{sec:fitting}). Since the effect of surface gravity and magnetic field are somewhat degenerate in broadening the lines, we kept the surface gravity fixed to the value obtained from photometry, and fitted for the best values of the magnetic field and temperature. We find that models with a magnetic field of 1.4~MG for M 39 and of 1.6~MG for ASCC 47 are an equally good fit to the Balmer lines as the non-magnetic models with the much higher surface gravities mentioned above. For both WDs, the magnetized fits result in slightly lower effective temperatures: $18,000\pm340$~K for M 39
and $110,000\pm4,000$~K for ASCC 47. The magnetized atmosphere models are quite simplistic, but the fit shows that the excessive broadening can be explained by the presence of the magnetic field and returns magnetic field values that are comparable with what we can infer from the Zeeman splitting of H$\alpha$.

\section{Initial-final mass relation}
\begin{figure}
    \centering
        \includegraphics[width=\columnwidth]{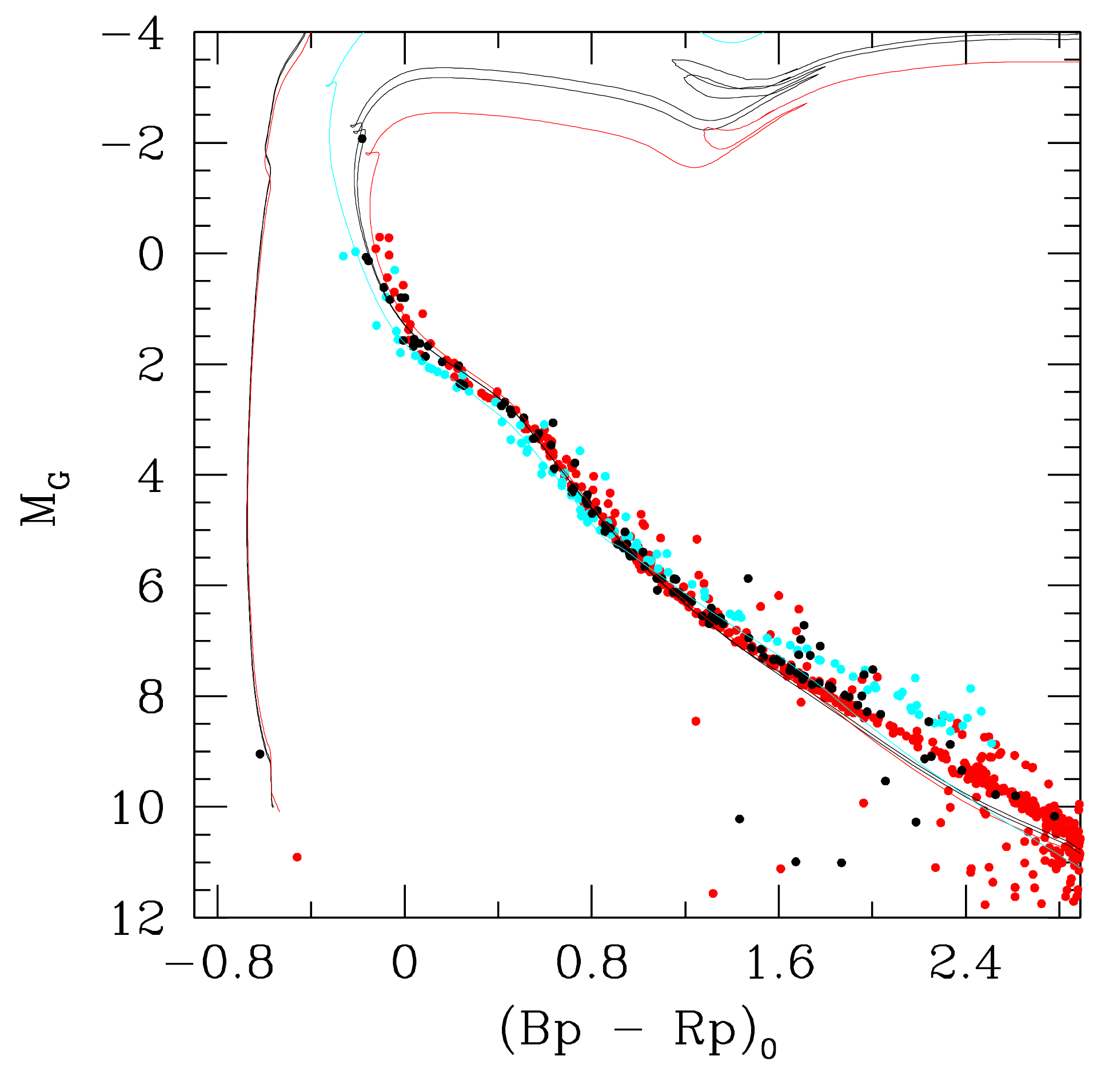}
     \caption{Gaia CMD of ASCC 47 (black) together with that of the Pleiades (135 Myrs, red) and NGC 2451B (44 Myrs, cyan) and MIST isochrones \citep{2016ApJS..222....8D,2016ApJ...823..102C,2011ApJS..192....3P,2013ApJS..208....4P,2015ApJS..220...15P} of the appropriate age for each cluster in the same colors. For ASCC 47, we plot two isochrones in black at 85 and 95 Myrs. The age of ASCC 47 is bracketed by these two clusters from consideration of both the turnoff region of the CMD and the pre-main sequence. Our best estimate for the age of ASCC 47 is 90$\pm20$ Myrs.}
    \label{fig:cmd1}
\end{figure}

By measuring the age of the cluster and the cooling age of the WDs, we can estimate the mass of the progenitor star. We fit isochrones generated with the PARSEC \citep{10.1111/j.1365-2966.2012.21948.x} and MIST models \citep{2016ApJS..222....8D,2016ApJ...823..102C,2011ApJS..192....3P,2013ApJS..208....4P,2015ApJS..220...15P} to estimate the age of each of the clusters (the results for ASCC are depicted in Fig.~\ref{fig:cmd1}), which is also the total age of the WD including its lifetime before becoming a WD. We use the techniques developed in \citet{2018AJ....156..165C}.  The age of the cluster ASCC 47 is $90 \pm 20$~Myr and that of M 39 is $280 \pm 20$~Myr. We estimated the age of M 47 to be $150 \pm 20$~Myr in Paper I. Both clusters are old enough that the age determination is not affected by stellar rotation \citep{2018AJ....156..165C}. We subtract the age of the WD determined from the white-dwarf cooling models \citep{2001ApJS..133..413B} from the cluster ages to determine the lifetime of the star that became the white dwarf and therefore its initial mass.

\begin{deluxetable*}{c|CCCC}
\tablecaption{Initial-final mass relation. \label{tab:imfrel}}
\tablehead{
\colhead{Cluster Name} &
\colhead{Cluster age} &
\colhead{WD mass} & \colhead{WD cooling age} & \colhead{Progenitor mass} \\
\colhead{} & \colhead{[Myr]} &
\colhead{$[\rm{M}_\odot]$} & \colhead{[Myr]} & \colhead{$[\rm{M}_\odot]$}
}
\startdata
Messier 39 & 280\pm20  & 0.95\pm0.02 &  175\pm 25 & 5.4\pm0.6 \\
ASCC 47  & 90\pm20 & 1.01\pm0.02 & 0.25 \pm 0.08 & $5.6\pm0.8$ \\
Messier 47 & 150\pm20 & 1.06\pm0.05 & 75\pm15 & 6.1\pm0.5 \\
\enddata
\end{deluxetable*}

By measuring the cooling ages of the three WDs, we find that they all have progenitors with zero-age-main-sequence masses above $5$~M$_\odot$ (see the values in Table~\ref{tab:imfrel}). Among the WDs found in the Gaia catalog to be unequivocally members of clusters, there are only two other objects with progenitor masses above 5~M$_\odot$: EGGR 25 (LB 1497) and GD 50 \citep{10.1111/j.1745-3933.2006.00240.x}, both in blue in Fig.~\ref{fig:ifmr} (for more details see Appendix~\ref{sec:selection}).
We re-analysed the spectra of these two objects using available spectroscopic data \citep{2011ApJ...743..138G} and we do not find any indication of a magnetic field. Therefore, among known cluster WDs with progenitor masses above 5~M$_\odot$, we find that three out of five are magnetic, a higher percentage than the 15-25\% found in general \citep{2004IAUS..224..879K,2007ApJ...654..499K,2019A&A...628A...1L} or of the $\sim10$\% found with magnetic fields exceeding 0.5 MG \citep{2018MNRAS.480.3942H,2019A&A...628A...1L}.  If the underlying fraction of magnetic white dwarfs in the clusters was similar to that within 20~pc (i.e. about 20\%), one would find three or more magnetic white dwarfs out of a sample of five only 5.8\% of the time. On the other hand, if we focus only on nearby white dwarfs with fields stronger than 1~MG (as we found in the clusters), the local fraction is even smaller (seven out of 123);  in this case, even allowing for uncertainties in the local frequency of white dwarfs with fields greater than 1 MG, the chance of finding three or more magnetic white dwarfs out of five would be less than 0.5\%.  Therefore, we can reject the hypothesis that the intermediate-mass progenitor white dwarfs examined in this paper have a similar fraction of stars with fields greater than 1 MG as the local sample at the two- or three-sigma level.  To find the confidence intervals on the frequency of magnetic WDs in open clusters we use a binomial distribution (which is appropriate for this situation), and a flat prior on the frequency because ours is the first measurement. We find the fraction of magnetic white dwarfs produced from intermediate mass stars to be 60\% and between 23\% and 90\% with 95.4\% confidence.

%{\bf **JSH: remove the following}
%Unfortunately, our sample is small and could be affected by small-number statistics. However, this result is already significant as finding such a high fraction of magnetic WDs is highly unlikely if one assumes the underlying distribution to be the same as for all WDs. If we assume the null hypothesis for the percentage of magnetic WDs to be the value found for all WDs in the 20 pc sample, namely 23/126, the p-value for finding three magnetic WDs by chance in a randomly selected group of five, in a Pearson's chi-squared test \citep{Pearson1900}, is less than 3\%. If we instead consider the distribution of WDs within 20 pc with fields exceeding 0.5~MG, 7/123, to be the null hypothesis, then the p-value goes down to less than 1\%. 
%.
\begin{figure}[tb]
    \centering
    \includegraphics[width=\columnwidth]{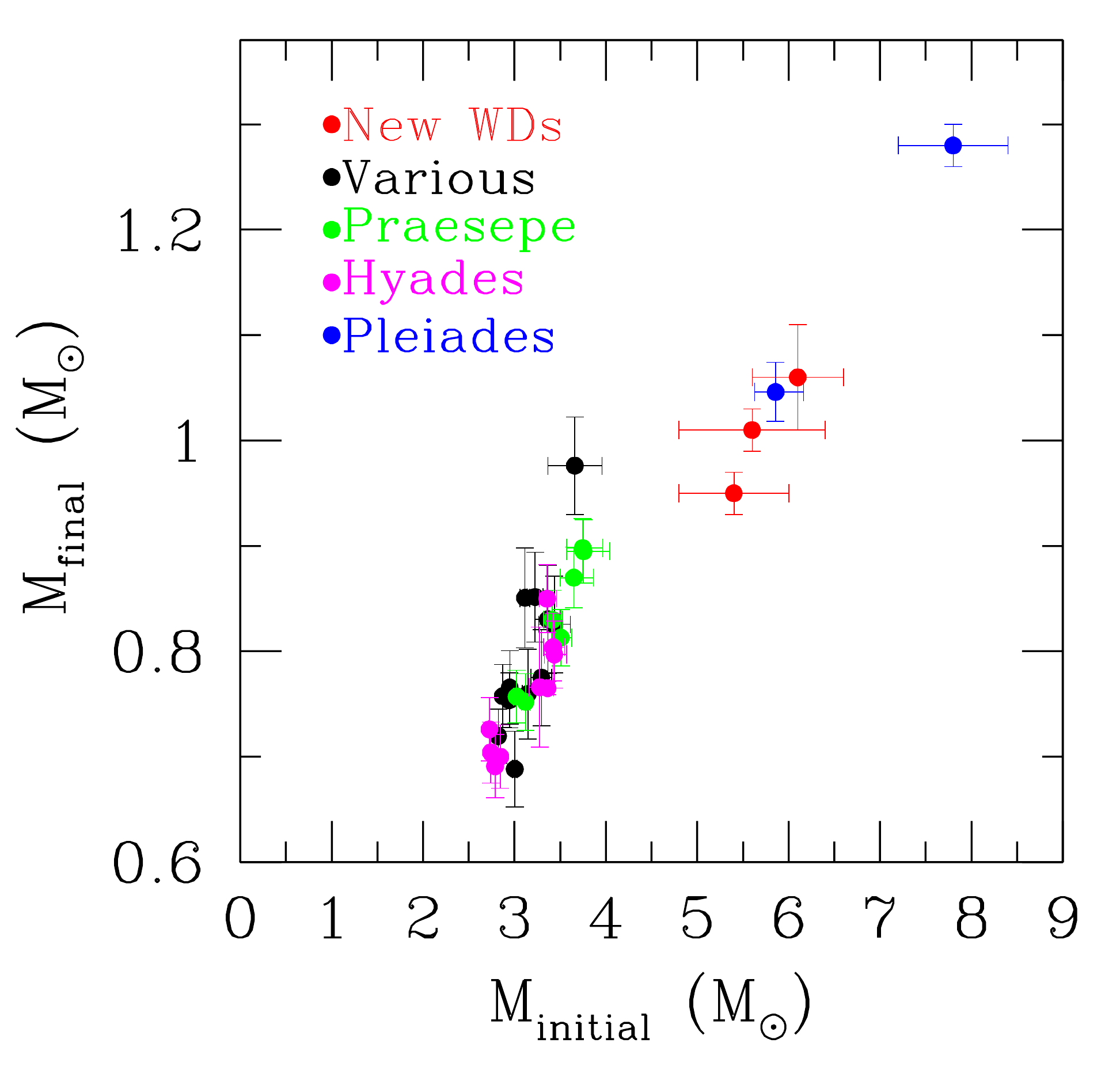}
    \caption{Initial-final mass relation (where the initial mass is the zero-age-main-sequence mass of the progenitor and the final mass is the mass of the WD) for cluster WDs that can be found in the Gaia catalog whose progenitors had masses above 2.5~M$_\odot$. Data from \citep{2018ApJ...866...21C,2018ApJ...861L..13G}. Error bars represent 1-$\sigma$ errors. The WDs reported in this paper and in Paper I are in red, other colors indicate rich clusters.} 
    \label{fig:ifmr}
\end{figure}

\section{Discussion}
The three WDs analyzed in this paper are high-fidelity cluster members, and their cooling ages, when compared to the ages of their respective parent clusters, are in agreement with a single-star evolution scenario and do not require any additional delay. Furthermore, if we go back in time to the moment in which the WDs were born, the mass of the stars in the parent clusters that were evolving into WDs were too high to 
%As mentioned above, the merger scenario cannot explain the presence of magnetic field in the three WDs because the parent clusters are too young to have produced 
produce the low-mass WDs that could merge and not explode in a supernova. A merger could have still occurred, and avoided a supernova, if binary interactions on the main-sequence or the post-main-sequence phases had led to a massive progenitor engulfing a lower mass star or substellar companion. It is hard to invoke such an evolutionary scenario for all three objects though, because the binary fraction in the three clusters is very low, as can be seen from the sparsely populated binary sequences in the clusters' CMDs (Fig.~\ref{fig:cmd}). Finding a high incidence of magnetic WDs in these young clusters therefore indicates the magnetic WDs can form directly from the evolution of single stars and do not necessarily result from the mergers of low-mass WDs. 

The WDs in M 39 and in ASCC 47 do not show any indication of a companion, as their color is consistent with the temperature inferred from the spectrum. The WD in M 47 does exhibit a red excess in its spectrum that could indicate either the presence of a disk or of a colder companion (see Paper I); however, the very narrow core of the Zeeman components in the spectrum rules out the presence of a companion in a close binary. Therefore, even though we cannot exclude that the magnetic field is the result of binary interactions after the WDs were born, the fossil origin of the field is the most likely explanation for the magnetism in these newly found WDs. The birth rates of magnetic Ap/Bp stars are much lower than the average occurrence of magnetic white dwarfs \citep{2004IAUS..224..879K}, and, therefore, magnetic main sequence stars cannot explain the even higher incidence of magnetic WDs that we found in this range of progenitor masses. 

Recent studies \citep{2015Sci...350..423F,2016Natur.529..364S,2016ApJ...824...14C} indicate that more than half of intermediate-mass stars host strong internal magnetic fields produced by powerful dynamos in their convective cores even though many do not exhibit strong surface fields like the magnetic Ap/Bp stars. In particular, they find that the presence and strength of the convectively driven magnetic dynamos in the cores show a strong dependence on stellar mass. Even more importantly, in stars with mass $M>3$~M$_\odot$, the convective core extends to a mass in the interior of the star that exceeds the mass of the WD descendant, allowing for the detection of the regions containing the high magnetic field in the final WD. This mechanism could explain why magnetic WDs are on average more massive and at the same time why we find a higher incidence of magnetic remnants from intermediate-mass stars.

\acknowledgments

We would like to thank Jim Fuller for useful discussions and the suggestion that the ASCC 47 white dwarf may be magnetic. The research was supported by NSERC Canada, the NSF, Compute Canada, a Burke Fellowship at Caltech and a Four-Year Fellowship at UBC. The research leading to these results has also received funding from the European Research Council under the European Union's Horizon 2020 research and innovation program n. 677706 (WD3D). This work has made use of data from the European Space Agency (ESA) mission
{\it Gaia} \\(https://www.cosmos.esa.int/gaia)
%, processed by the {\it Gaia} Data Processing and Analysis Consortium (DPAC, https://www.cosmos.esa.int/web/gaia/dpac/consortium). Funding for the DPAC has been provided by national institutions, in particular the institutions participating in the {\it Gaia} Multilateral Agreement. This work 
and is based on observations obtained under program IDs GS-2019B-FT-104 (ASCC 47),   GS-2018B-FT-108 (M 47) and GN-2019A-FT-214  (M 39) at the international Gemini Observatory.
%, a program of NSF’s OIR Lab, which is managed by the Association of Universities for Research in Astronomy (AURA) under a cooperative agreement with the National Science Foundation on behalf of the Gemini Observatory partnership: the National Science Foundation (United States), National Research Council (Canada), Agencia Nacional de Investigaci\'{o}n y Desarrollo (Chile), Ministerio de Ciencia, Tecnolog\'{i}a e Innovaci\'{o}n (Argentina), Minist\'{e}rio da Ci\^{e}ncia, Tecnologia, Inova\c{c}\~{o}es e Comunica\c{c}\~{o}es (Brazil), and Korea Astronomy and Space Science Institute (Republic of Korea). 
This work is also based on data products from observations made with ESO Telescopes at the La Silla Paranal Observatory under program ID 177.D-3023, as part of the VST Photometric H{alpha} Survey of the Southern Galactic Plane and Bulge (VPHAS+, www.vphas.eu), and on data from the Pan-STARRS1 Survey (PS1) and the PS1 public science archive. 
%have been made possible through contributions by the Institute for Astronomy, the University of Hawaii, the Pan-STARRS Project Office, the Max-Planck Society and its participating institutes, the Max Planck Institute for Astronomy, Heidelberg and the Max Planck Institute for Extraterrestrial Physics, Garching, The Johns Hopkins University, Durham University, the University of Edinburgh, the Queen's University Belfast, the Harvard-Smithsonian Center for Astrophysics, the Las Cumbres Observatory Global Telescope Network Incorporated, the National Central University of Taiwan, the Space Telescope Science Institute, the National Aeronautics and Space Administration under Grant No. NNX08AR22G issued through the Planetary Science Division of the NASA Science Mission Directorate, the National Science Foundation Grant No. AST-1238877, the University of Maryland, Eotvos Lorand University (ELTE), the Los Alamos National Laboratory, and the Gordon and Betty Moore Foundation. 
We used the Montreal White Dwarf website (http://www.astro.umontreal.ca/~bergeron/CoolingModels), the PARSEC stellar models (http://stev.oapd.inaf.it/cgi-bin/cmd), the MIST stellar models \\ (http://waps.cfa.harvard.edu/MIST) and the VizieR catalog access tool, CDS, Strasbourg, France (DOI : 10.26093/cds/vizier). 
%The original description of the VizieR service was published in 2000, A\&AS 143, 23.


\vspace{5mm}
\facilities{Gaia, Gemini Observatory(GMOS), VLT(VPHAS+ Survey), Pan-STARRS(PS1 Survey)}


\software{IRAF \citep{1986SPIE..627..733T}
          }



\appendix

\section{Cluster membership}
\label{sec:membership}

Stellar clusters are concentrations of stars in the phase space of position and velocity.  For most stars in the catalog, Gaia gives five out of six of the phase-space coordinates: velocity across the sky (proper motion), position on the sky and distance. We identified the member stars of the two clusters by extracting all of the stars within a given radius on the sky from the centre of the cluster (cone search of 2.5 and 1.5 degrees for ASCC 47 and M 39 respectively). An initial photometric cut was done, discarding stars with a colour excess factor above a threshold value (1.5 for ASCC 47 and 1.8 for M 39) in the Gaia parameter $B_p-R_p$. This reduced false WD candidates and contamination from non-member field stars in the final sample. Among the selected stars, we located a concentration in proper motion space and kept stars within four standard deviations of the cluster's mean proper motion. %The mean and deviation were derived by fitting a 2-dimensional gaussian function to a 2-dimensional histogram of proper motion.
We then identified a mean parallax value for the cluster and assumed stars within four standard deviations of this central peak were members, as depicted in the left panels of Fig.~\ref{fig:cluster_and_WD}. In both clusters, we identified WDs which are indicated by red bars and crosses in Fig~\ref{fig:cluster_and_WD}. The photometric and kinematic properties of the cluster and the WD are given in Tables~\ref{tab:photometry} and~\ref{tab:astrometry}. The reddening values that we use are the average of the Gaia reddening values of cluster members (Gaia only provides reddening values for bright stars). For a discussion on the reddening of the WDs, see Appendix~\ref{sec:phot}.

\begin{table}
\centering
 \caption{WD and Cluster Photometry}
\scriptsize
\medskip
\begin{tabular}{c|cccccc}
\hline
Cluster & Gaia WD ID & G$_{obs}$  & (BP$-$RP)$_{obs}$ & E(BP - RP)$_{cluster}$ & G$_0$ & (BP$-$RP)$_0$\\
\hline
Messier 47  & 3029912407273360512 & $19.796\pm0.006$ & $-0.134\pm0.106$& $0.08\pm0.03$ & $11.12\pm0.03$ & $-0.22\pm0.11$\\
Messier 39 & 2170776080281869056& $19.193\pm0.003$& $-0.179\pm0.050$& $0.15\pm0.04$ & $11.52\pm0.04$& $-0.33\pm0.06$\\
ASCC 47 & 5529347562661865088 & $18.714\pm0.003$& $-0.509\pm0.015$& $0.11\pm0.04$ & $9.04\pm0.03$ & $-0.62\pm0.04$\\
\hline
\end{tabular}
\label{tab:photometry}
\end{table}

\begin{table}
\centering
 \caption{WD and Cluster Astrometry}
\scriptsize
\medskip
\begin{tabular}{c|cccccccc}
\hline
Cluster & cluster parallax & WD parallax & cluster $\mu_{RA}$ & cluster $\mu_{dec}$ & WD $\mu_{RA}$ & WD $\mu_{dec}$\\
\hline
Messier 47 & $2.072\pm0.096$ & $2.313\pm0.608$ & $-7.047\pm0.193$ & $0.977\pm0.177$ & $-7.174\pm0.873$ & $1.266\pm0.868$ \\
Messier 39 & $3.350\pm0.042$ & $3.309\pm0.277$ & $-7.472\pm0.131$ & $-19.848\pm0.152$ & $-6.631\pm0.462$ & $-20.181\pm0.456$ \\
ASCC 47    & $1.273\pm0.046$   & $1.299\pm0.214$ & $-7.839\pm0.355$ & $5.562\pm0.329$ & $-7.923\pm0.309$ &$5.377\pm0.385$& \\
\hline
\end{tabular}
\label{tab:astrometry}
\end{table}

\section{Spectral Fitting} 
\label{sec:fitting}

The analysis of the spectroscopic data consists of several steps that we also simulate to determine our measurement uncertainties. We employ atmospheric models developed by Gianninas et al. 2010 \citep{2010ApJ...720..581G} for ASCC 47 and by Tremblay et al. 2011 \citep{2011ApJ...730..128T} for M~39. In both sets of models, the hydrogen atmosphere is computed without the assumption of local thermodynamic equilibrium; the main difference is that in the former, the composition of the atmosphere includes carbon, nitrogen, and oxygen at solar abundance ratios, while the latter are made of pure hydrogen. The addition of metals in the atmosphere is important for very hot WDs, where metal levitation in the intense radiation field can change the shape of the Balmer lines \citep{2010ApJ...720..581G}.  First, we normalize the flux by fitting a tenth-order polynomial in wavelength to the spectra, avoiding the Balmer lines, to account for potential errors in the broadband calibration. We use the same normalization procedure for both the models and the observed spectrum. We convolve the models with the instrumental Gaussian profile (5~\AA). The fit is then carried out for the four Balmer lines H$\beta$, H$\gamma$, H$\delta$ and H$\epsilon$ only, using a Levenberg-Marquardt method. In the case of the non-magnetic models, our free parameters are: $\log g$, $T_{\rm{eff}}$, a redshift common to all lines and a zero point shift for each line.  For the magnetic case, we take the value of $\log g$ to be fixed to the value determined by the photometry and allow the magnetic field to vary, together with the other parameters. 

\begin{figure}
    \centering
    \includegraphics[width=0.245\textwidth]{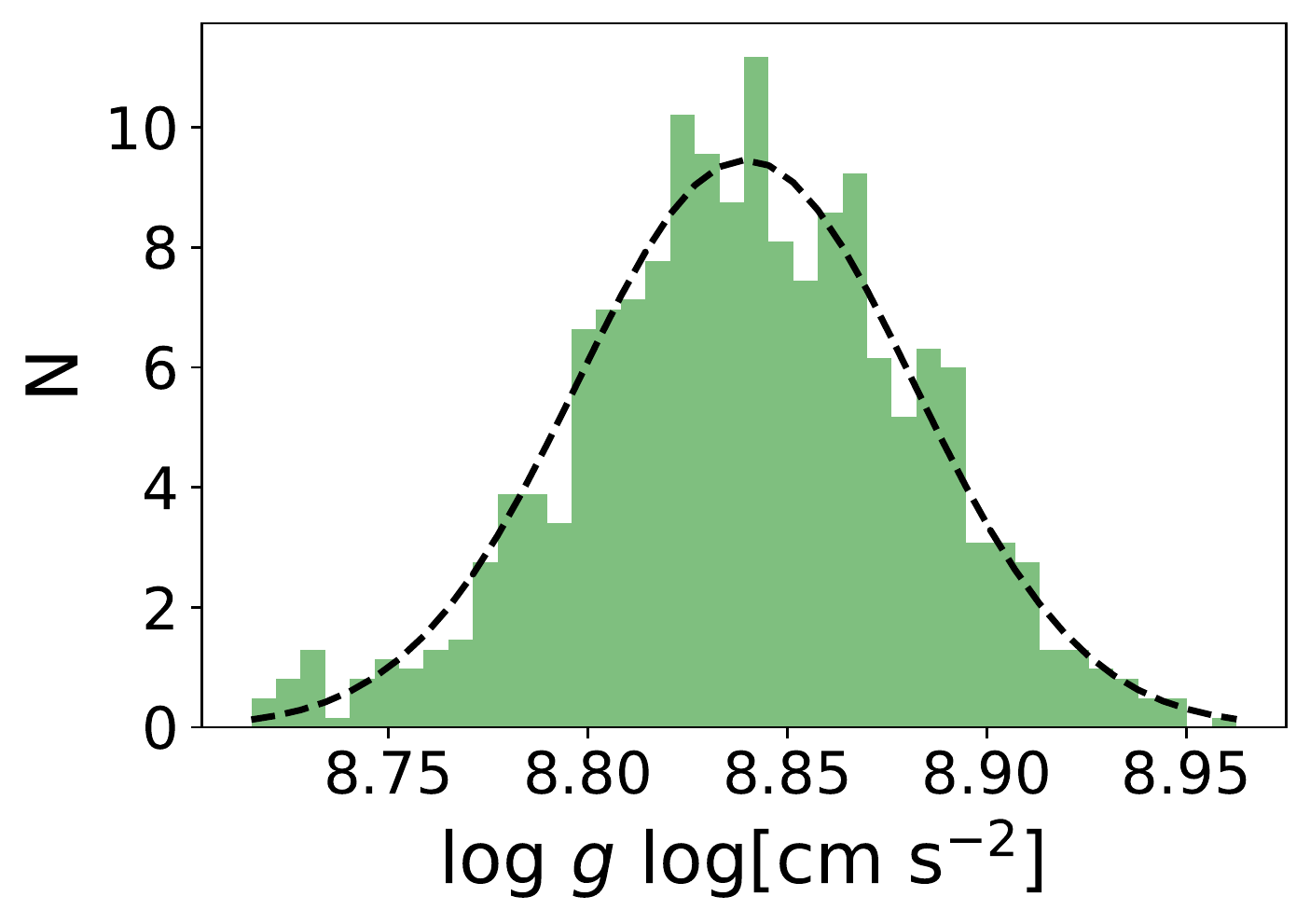}
    \includegraphics[width=0.245\textwidth]{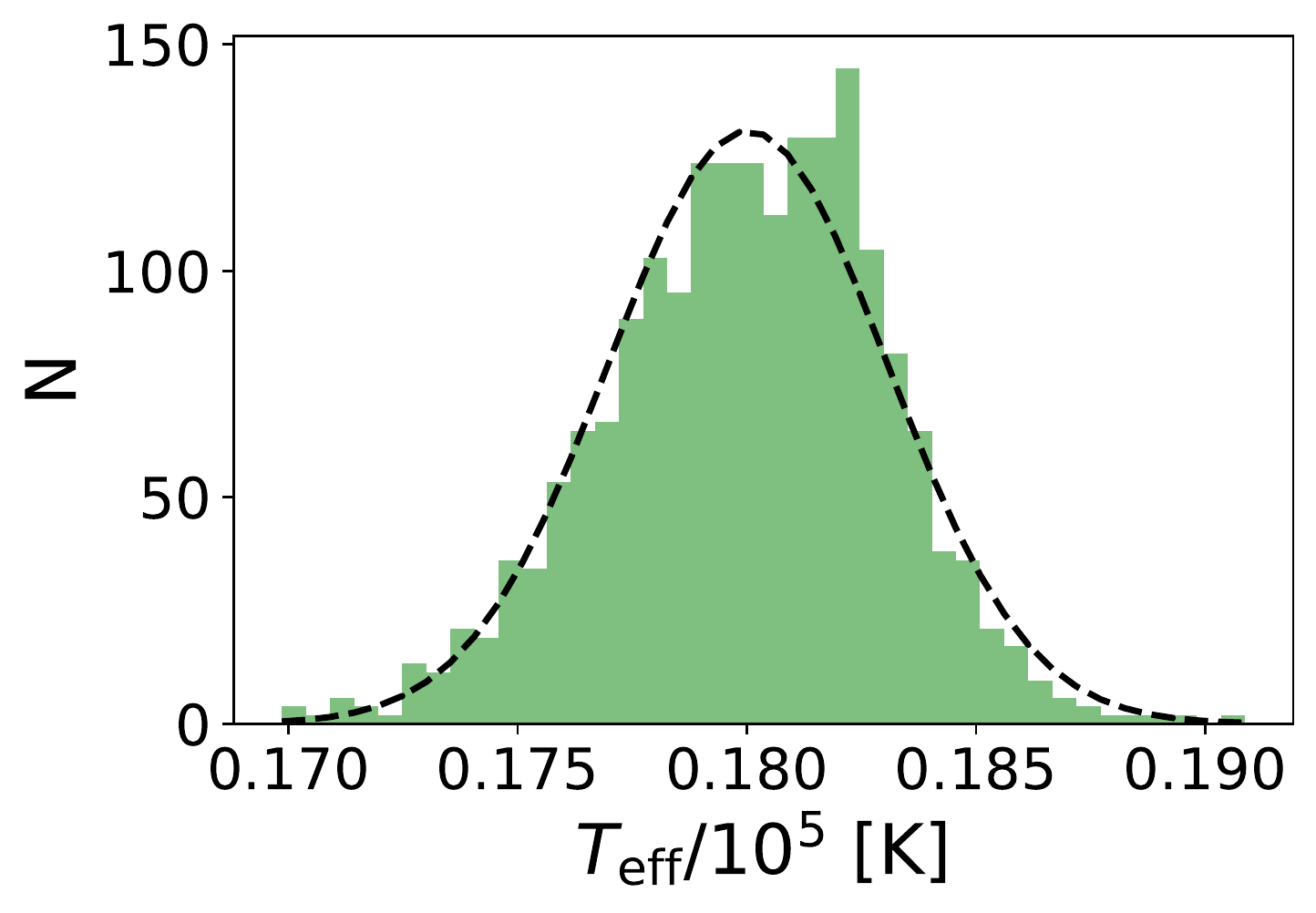}
    \includegraphics[width=0.245\textwidth]{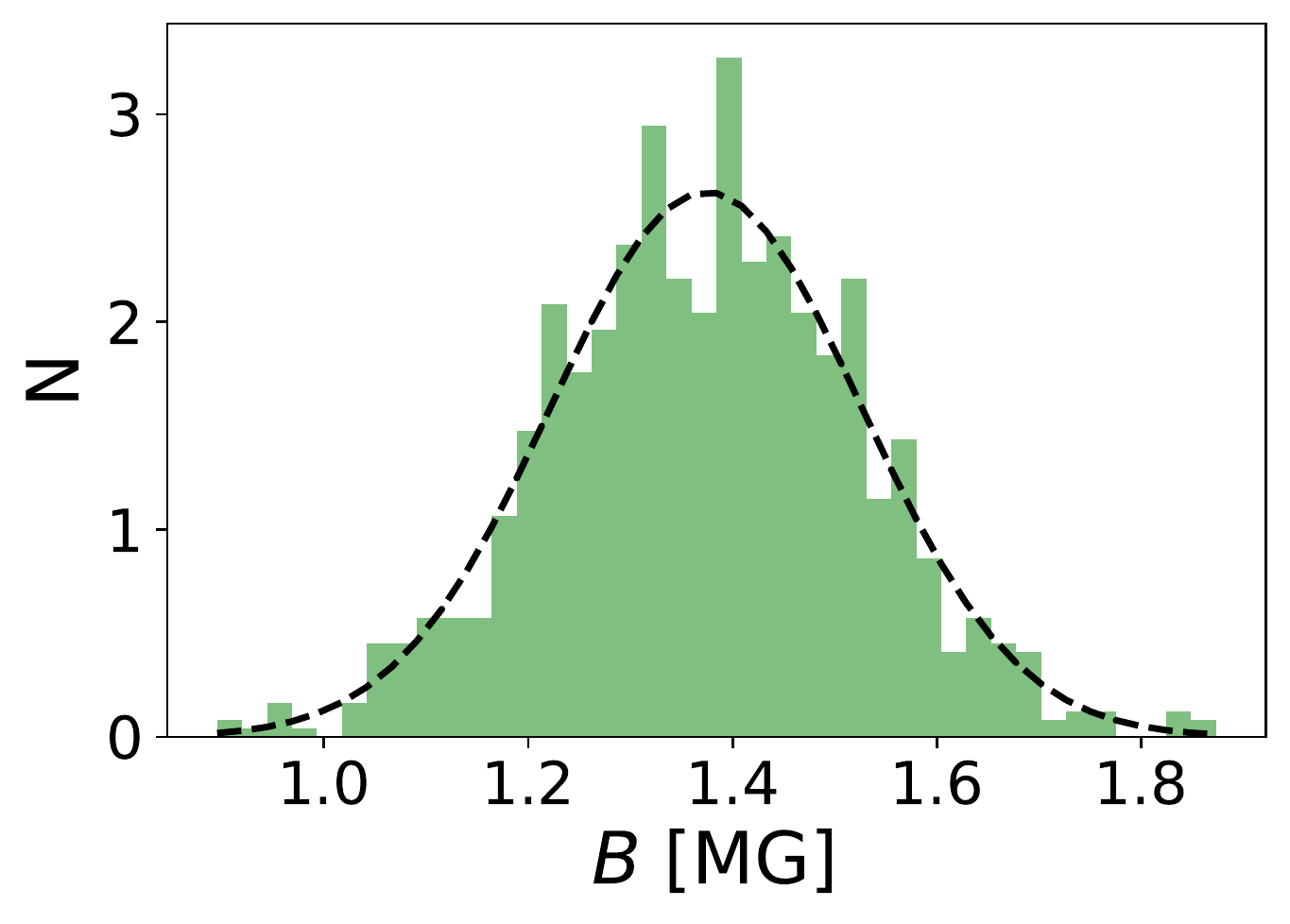}
    \includegraphics[width=0.245\textwidth]{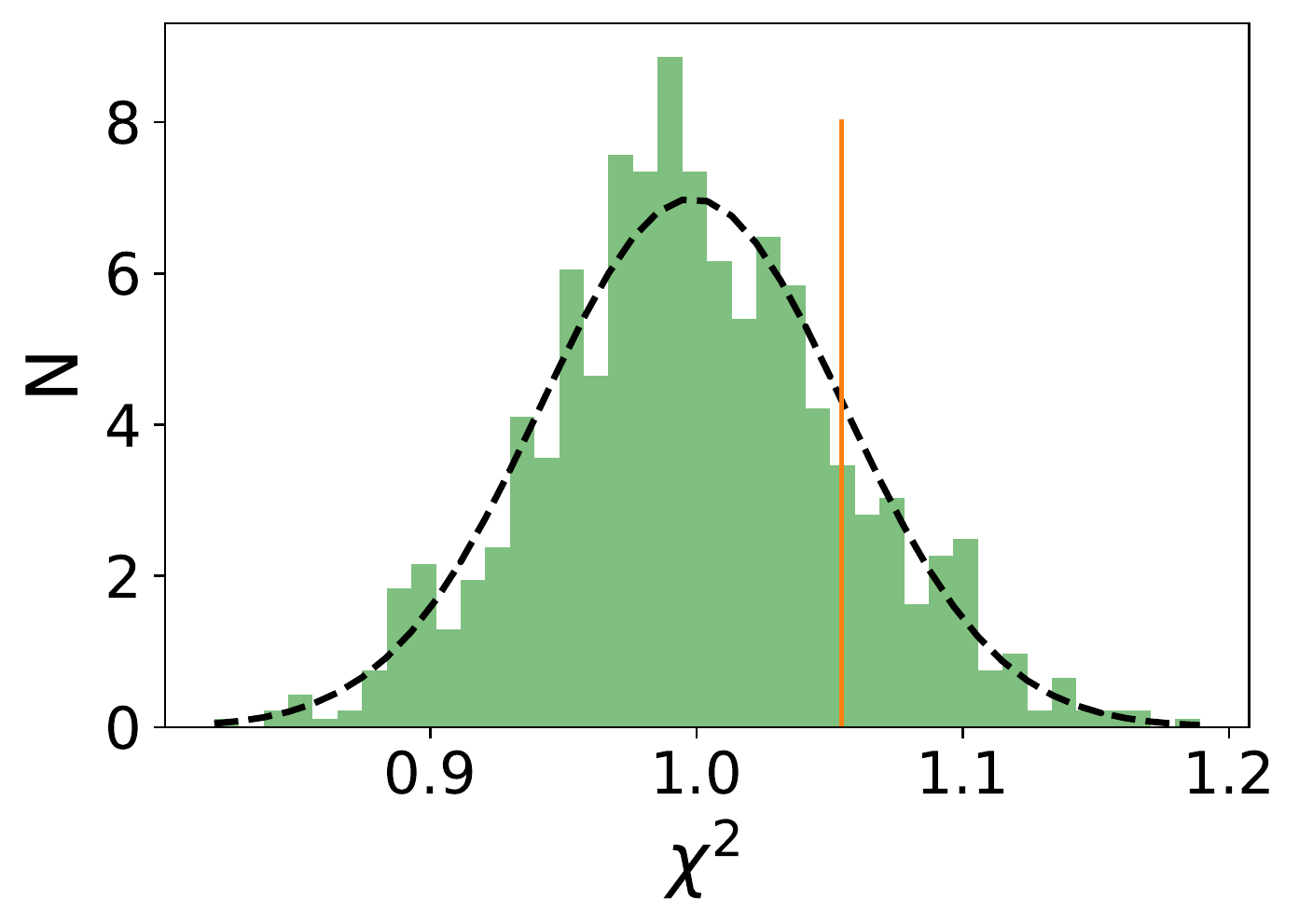}
    \includegraphics[width=0.245\textwidth]{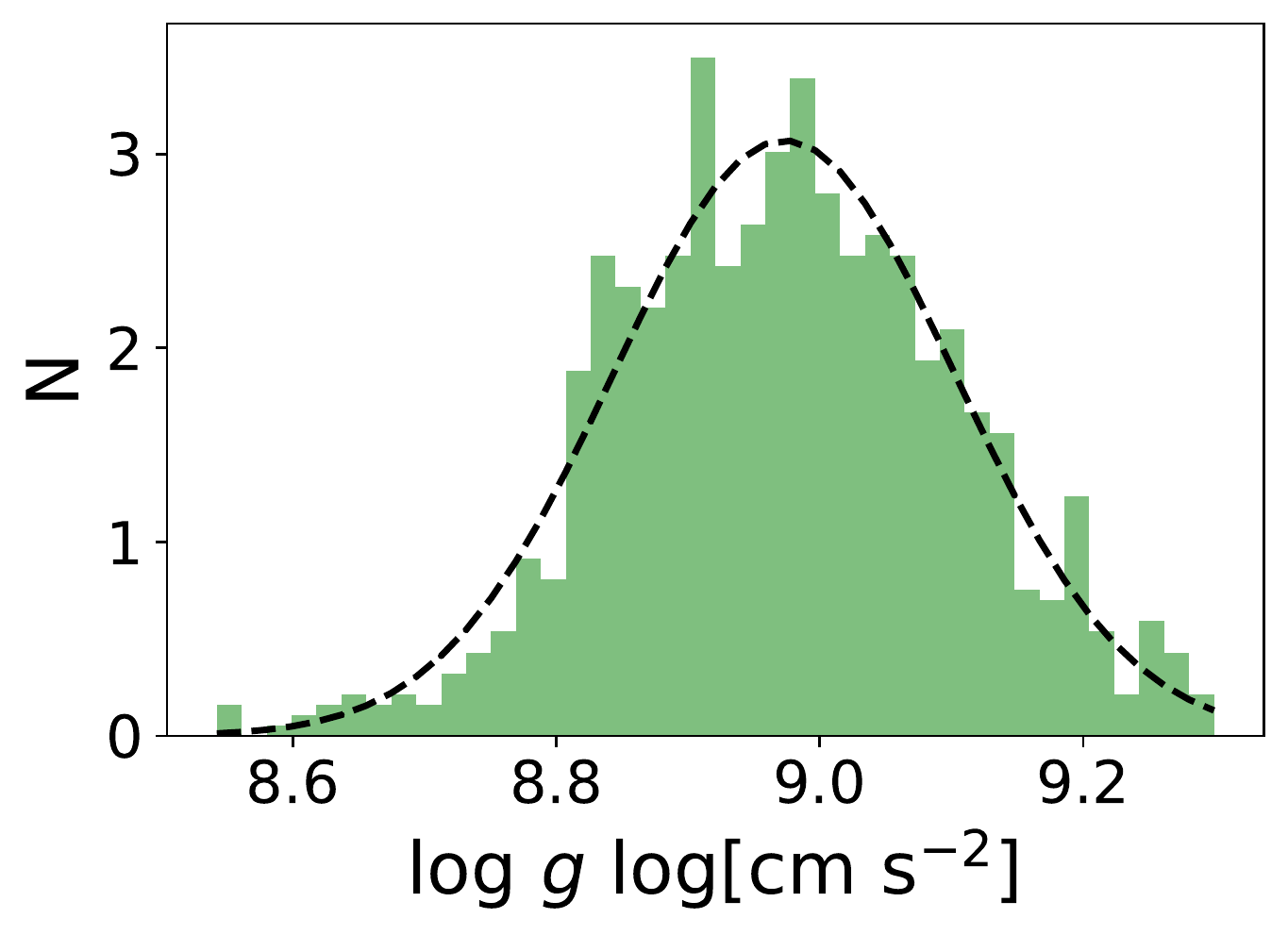}
    \includegraphics[width=0.245\textwidth]{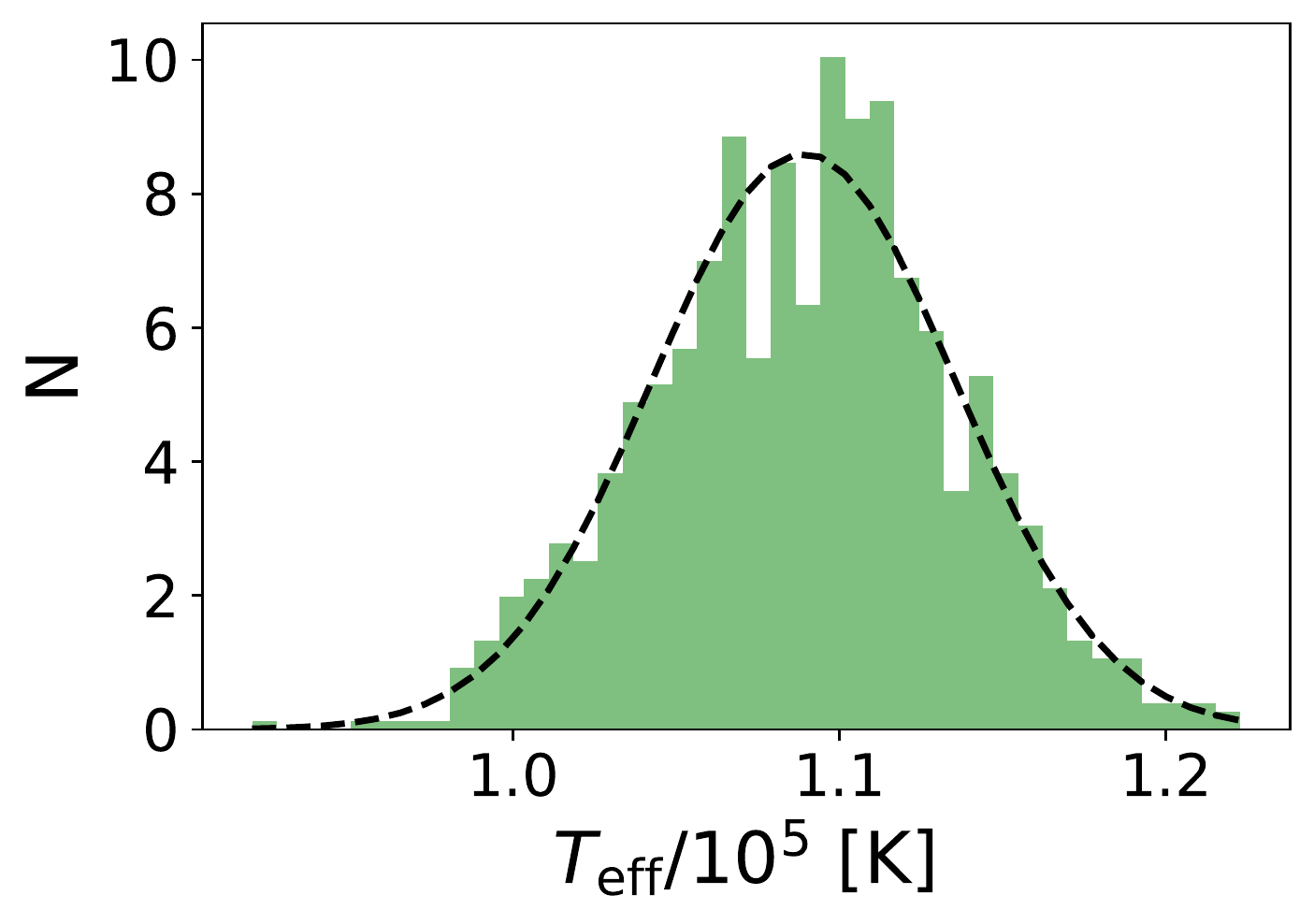}
    \includegraphics[width=0.245\textwidth]{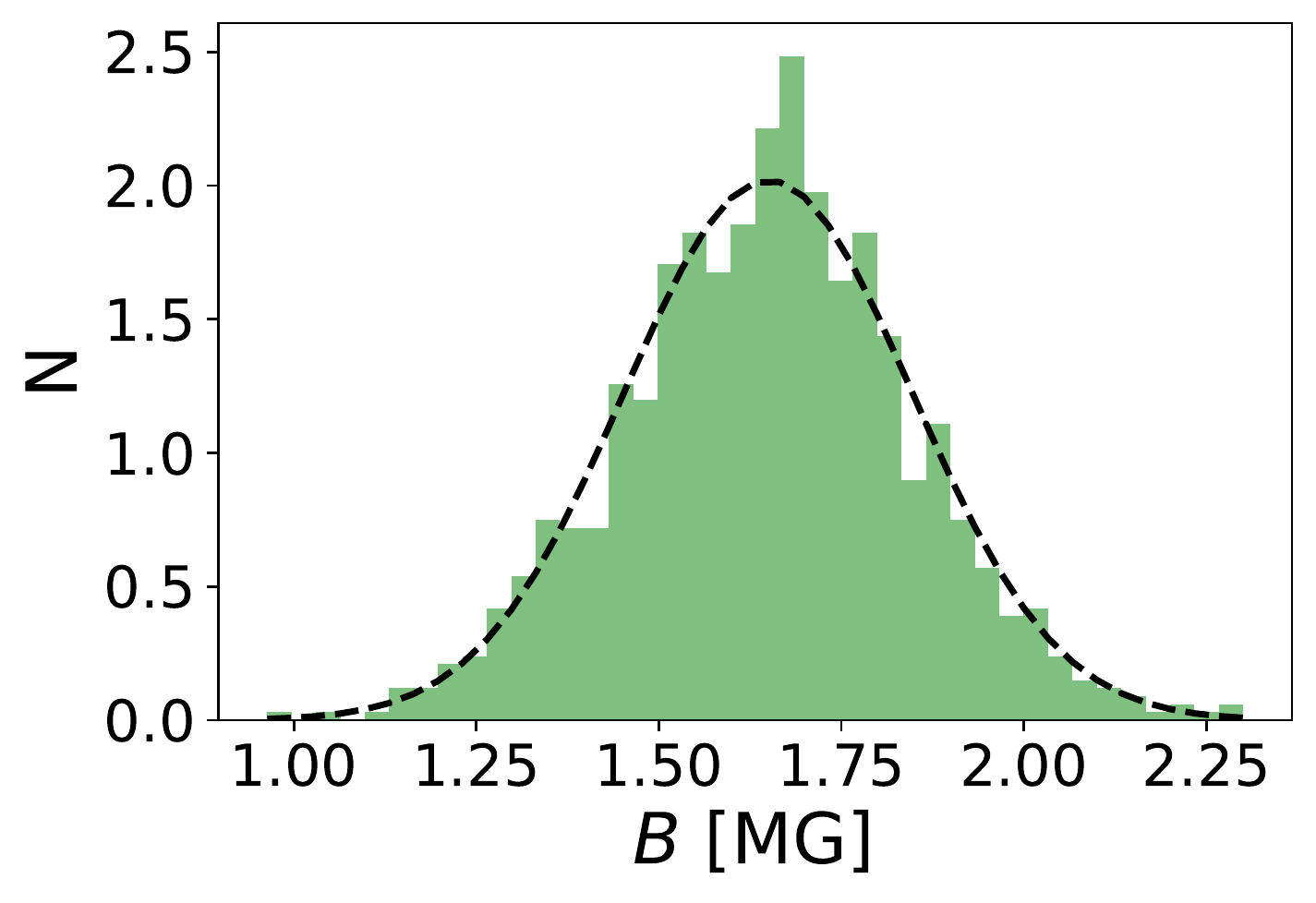}
    \includegraphics[width=0.245\textwidth]{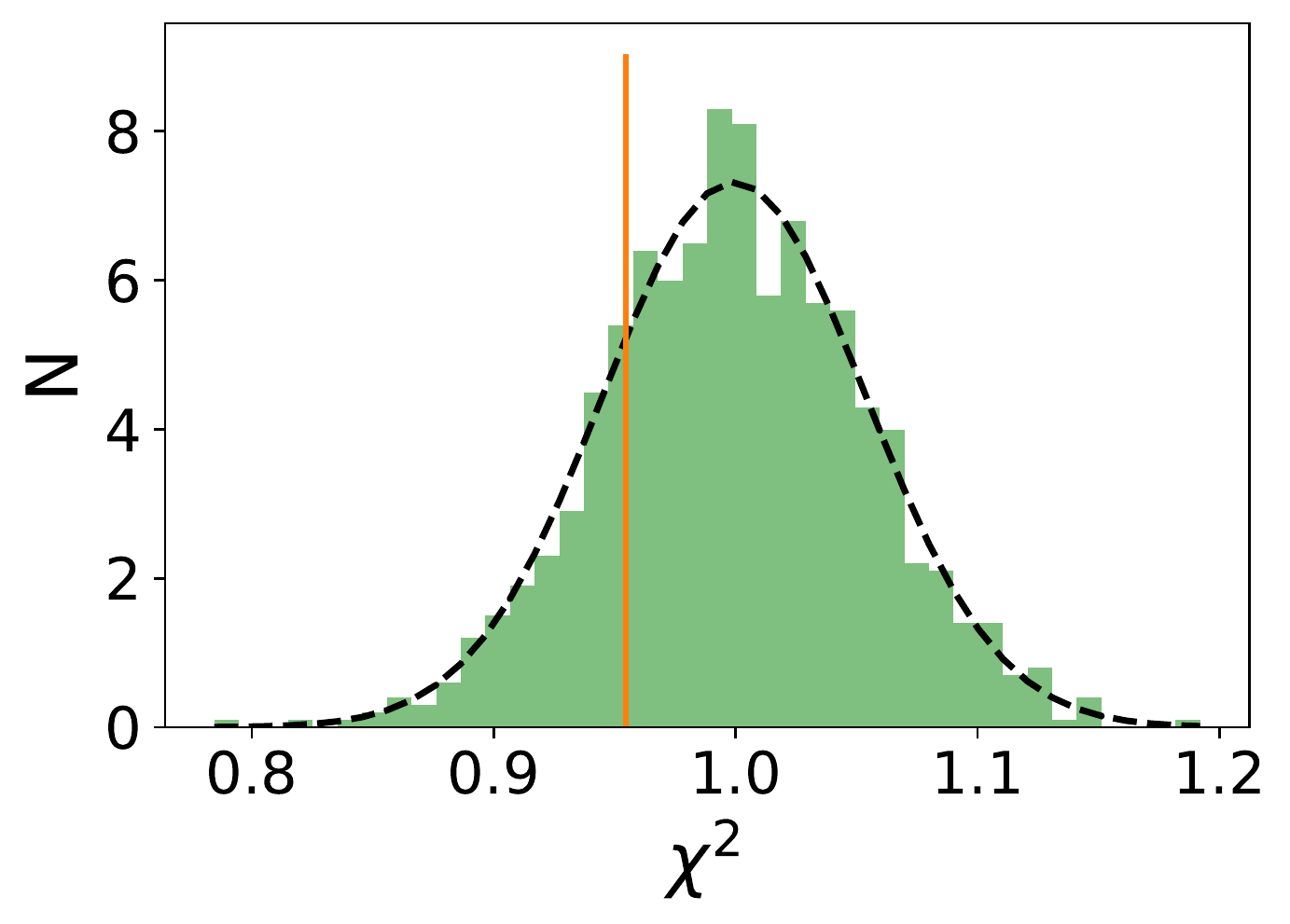}
    \caption{The results of our MC simulations for a total of 1,000 realizations for M 39 (upper row) and ASCC 47 (lower row). The distributions for the retrieved parameters are Gaussian centered on the input parameters. The first plot in each row shows the distribution of $\log g$ retrieved from the non-magnetized models; the second and the third show $T_{\rm{eff}}$ and $B$ retrieved from the magnetized models; the last plot shows in green the distribution of $\chi^2$ for the non-magnetic models and the orange line shows the $\chi^2$ value for the best magnetic model.}
    \label{fig:Sim}
\end{figure}

In order to understand the uncertainties in our fitting procedure, we simulate the fitting process. We take a model spectrum with the same parameters as in our best fit. To add low-frequency noise that could simulate calibration artefacts, we multiply the continuum of the spectrum with the same polynomial that we use to normalize the data. We add Gaussian noise to the resulting simulated spectrum. Finally, because our spectral data is oversampled, we smooth the simulated spectrum so that it has similar noise properties to the observations.  We generate and fit an ensemble of Monte Carlo (MC) spectral observations to determine the uncertainties in our parameter fits and the significance of differences in the quality of the fits between the magnetic and non-magnetic spectral models to our data. We check that the noise properties are reproduced correctly in the simulation by making sure that the average $\chi^2$ is the same as the $\chi^2$ obtained for the best-fitting model to the real data. A sample of our simulations is shown in Fig.~\ref{fig:Sim}: the histograms show the parameters retrieved in the MC simulation, and we find that that the distributions are Gaussian with the mean centered on the input value; the upper plots are for M 39 and the lower ones are for ASCC 47.

We employ simple magnetic models to fit the Balmer lines as well. As we are considering magnetic fields below 10~MG, the effect of magnetic field on the atomic structure of hydrogen reduces to the linear Zeeman effect, for which the degeneracy in the electron's energy levels in the magnetic quantum number $m_l$ is lifted, and all energy levels are shifted by an amount $\frac{1}{2} m_l h \omega_c$, where $\omega_c=eB/(m_e c)$ is the cyclotron frequency, $B$ is the magnetic field and $m_e$ is the mass of the electron. Balmer lines are therefore split into three components, with the central component being at the same wavelength as the zero-field line and a blueshifted ($\Delta m_l = +1$) and a redshifted ($\Delta m_l = -1$) component. The separation in energy is given by $\Delta E = \pm \frac{1}{2} h \omega_c = \pm 5.79\times10^{-3} \left(\frac{B}{1~\rm{MG}}\right)$~eV, and therefore the shifted components are centred at the wavelengths $\lambda_0 \pm \Delta \lambda_\pm = \lambda_0\left(1\mp\lambda_0\frac{\Delta E}{ch}\right)^{-1}$, where $\lambda_0$ is the zero-field wavelength. From this linear relation between field and wavelength separation, we can infer the magnetic field strength by analyzing the splitting in the H$\alpha$ line. For both our DA WDs, the separation is about 20~\AA, and therefore the field strength is about 1~MG.  In this linear regime, pressure broadening is dominant, and it is safe to assume, as a first approximation, that each of the three Zeeman components are Stark broadened as in the zero-field case \citep{1998MNRAS.299L...1F}. Starting from the models by Gianninas et al. \citep{2010ApJ...720..581G} for ASCC 47 and the models by Tremblay et al. \citep{2011ApJ...730..128T} for M 39, we created magnetic models with the same continuum as the non-magnetic ones, and with the Balmer lines split in the three components, with each component Stark broadened as the zero-field line. The total  equivalent width of each Balmer line is the same as for the zero-field case, and the central component contributes 50\% of the flux, while the redshifted and blueshifted components contribute 25\% each. Our models do not include any consideration on the structure of the magnetic field, as they assume the strength of the field to be the same over the surface of the stars. Using more sophisticated models, as the ones presented in \citep{1979MNRAS.188..165W,1981MNRAS.196...23M,1997MNRAS.292..205F}, would not provide more information as the signal-to-noise and resolution in our spectra would not allow us to distinguish among possible field structures. Using the fitting procedure and simulations outlined above, we find that our best magnetic models fit the Balmer lines of the objects with a $\chi^2$ that is within one sigma of the best non-magnetic model fit, and therefore they are statistically equally good fits. The comparison is shown in the right-most plots of Fig.~\ref{fig:Sim}: the histogram in green shows the distribution of $\chi^2$ for the non-magnetic models and the orange line shows the $\chi^2$ value found for the best magnetic model: for ASCC 47, the best magnetic model fits the Balmer lines better than the best fitting non-magnetic model, while for M 39 the fit is slightly worse, but both are within one sigma of the distribution. 

\section{Photometric fitting} 
\label{sec:phot}

\begin{figure}
    \centering
    \includegraphics[width=0.48\textwidth]{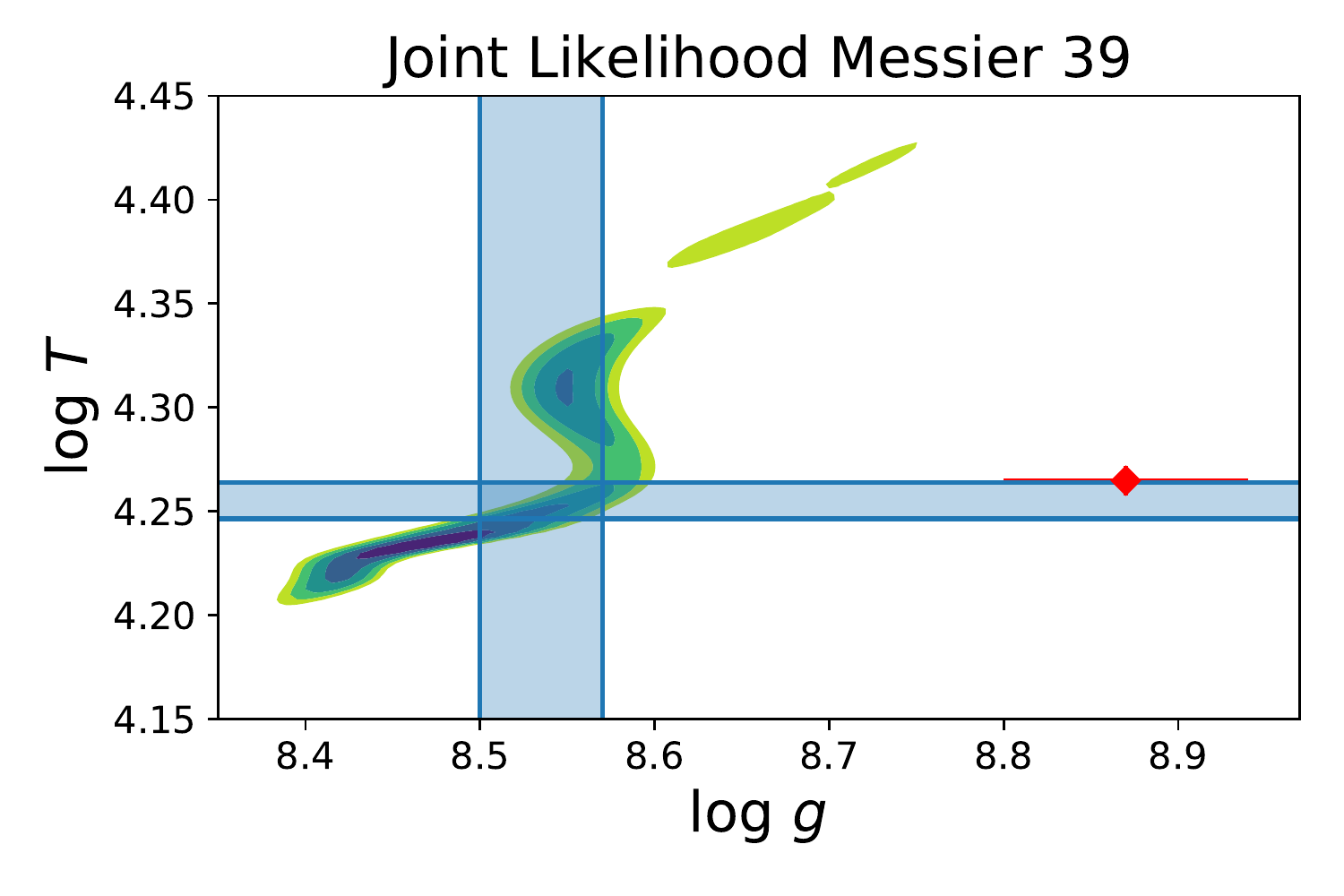}
    \includegraphics[width=0.48\textwidth]{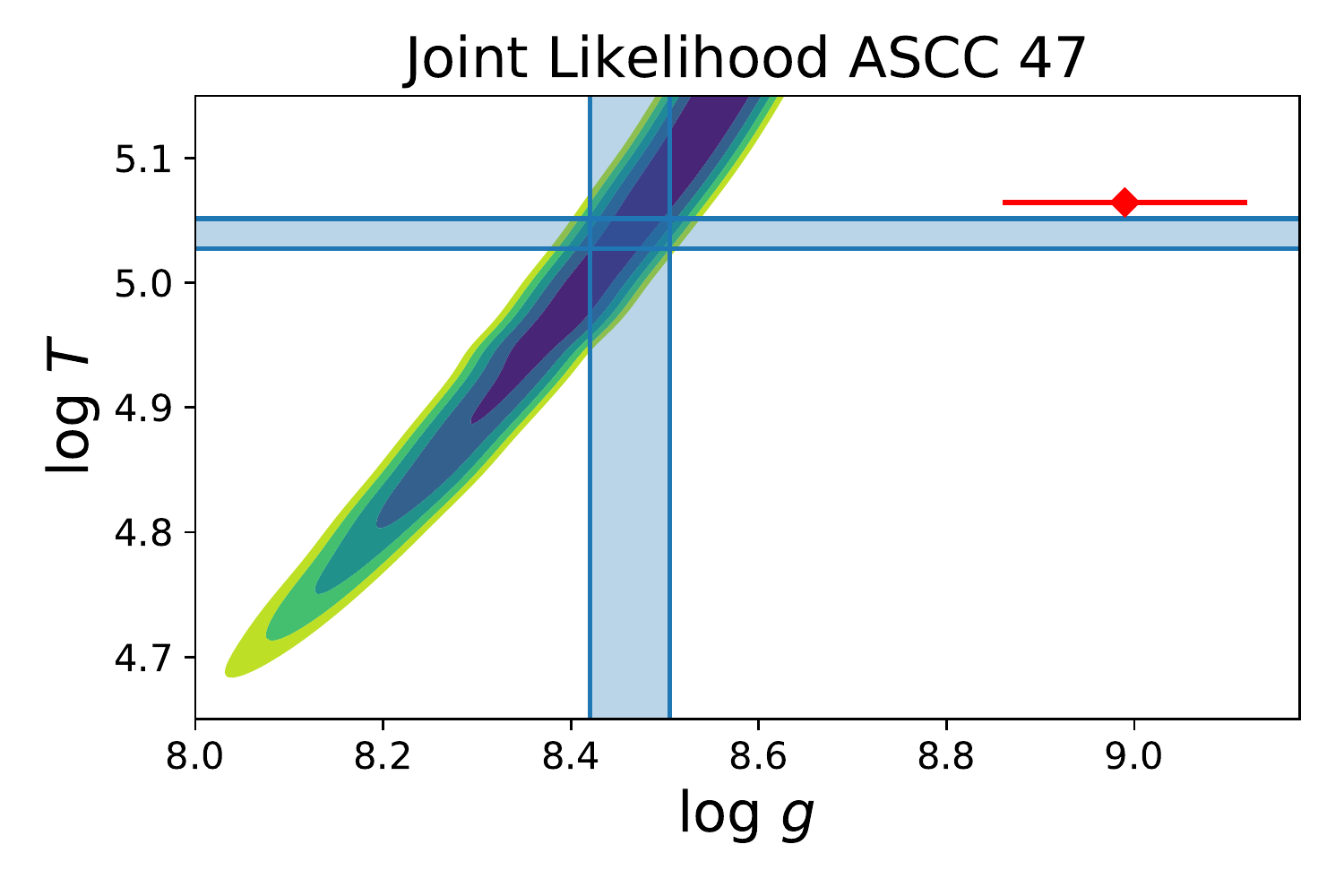}
    \caption{The joint likelihoods for the fits of the photometric data for the WDs in Messier 39 and ASCC 47 to white-dwarf models.  The values of the surface gravity and temperature obtained from the spectroscopic fits using non-magnetic models are depicted with red errorbars.  The corresponding best-fit temperature range from spectroscopic fits with magnetic models are shown with a horizontal band.   These constraints  combined with the photometric fits yield a confidence interval on the surface gravity depicted by the vertical band.}
    \label{fig:joint-likelihoods}
\end{figure}

To estimate the masses, radii and ages of the WDs from the photometry, we use synthetic photometric models \citep{2001ApJS..133..413B,2006AJ....132.1221H,2006ApJ...651L.137K,2011ApJ...730..128T,2018ApJ...863..184B} combined with Gaia \citep{2018A&A...616A...1G} and Pan-STARRS \citep{2016arXiv161205560C} data for the M 39 WD and with Gaia and VPHAS+ photometry \citep{2014MNRAS.440.2036D} for the ASCC 47 WD and the Gaia estimates of the distances to the two clusters.  For both WDs we used photometry in the following bands: $G$, $B_p$ and $R_p$ from Gaia, $g$, $r$ and $i$ from Pan-STARRS and VPHAS+. The photometry was insufficient to determine the interstellar reddening to the white dwarf, so we marginalized our fitting procedure over the measured reddening values for the stars in each cluster from Gaia and applied reddening corrections \citep{1989ApJ...345..245C,1994ApJ...422..158O} to our synthetic photometry models.  Fig.~\ref{fig:joint-likelihoods} depicts the one through five-sigma confidence regions for the fits to the photometric data.  The width of the confidence region is mainly determined by the uncertainty in the distances to the clusters.  The vertical span of the confidence region results from the uncertainty in the interstellar absorption toward the WDs. On both plots, the temperature and surface gravity determined by fitting non-magnetic spectral models to the spectroscopic data are depicted with red error-bars. For neither WD is the photometry consistent with the surface gravity and temperature measured in this way. Because of the uncertainty in the interstellar reddening, the photometry cannot constrain the temperature on its own.  However, the strength of the spectral lines depends strongly on the temperature of the star (and is independent of the reddening), so we can use the spectroscopy to constrain the temperature and find a value of the surface gravity and temperature that is consistent with the photometry and spectroscopy as long as the white dwarf has a magnetic field on the order of a million Gauss (the difference in temperature between fitting a magnetic or a non-magnetic model is small, within the error bars, see main text).  The photometric fitting also yields an estimate of the ages of the two WDs: $250\pm 80$~kyr for ASCC 47 and $175 \pm 25$~Myr for M 39.  When we combine these estimates for the ages of the WDs with the cluster ages, we can estimate the initial masses of the stars that became the two WDs as tabulated in Table~1 in the main text.

\section{Gaia cluster white dwarfs} 
\label{sec:selection}

We are interested in the total number of cluster white dwarfs with progenitor masses above $5$~M$_\odot$ whose cluster membership can be confirmed through Gaia astrometry. In order to select the sample of cluster white dwarfs that is shown in Fig.~\ref{fig:ifmr} in the main text, we selected all the white dwarfs known to be cluster members that are also part of the Gaia catalog and checked their membership using Gaia astrometry. Some known WDs that are likely associated with clusters are too faint to be in Gaia and therefore we did not include them. In the progenitor mass range above $5$~M$_\odot$, we found 4 white dwarfs in Gaia that were previously associated with clusters. We retained two of them, GD 50 and EGGR 25. EGGR 25 respects our standards for cluster membership in the Pleiades, while GD 50, the most massive object in our sample, is located well outside of the Pleiades cluster boundary. A recent study showed that GD 50 is very likely a member of the AB Doradus moving group \citep{2018ApJ...861L..13G}, which is virtually coeval with the Pleiades, and therefore we can still include the white dwarf in our sample as we can infer a progenitor mass. The remaining two are a white dwarf in NGC 1039 \citep[WD 17][]{2008AJ....135.2163R} and PG~0136+251, historically associated with the Pleiades \citep{10.1111/j.1745-3933.2006.00240.x}. Both the parallax and proper motion of the first object rule out its membership in NGC 1039, and we therefore excluded this WD from our sample. We performed a similar analysis as in \citet{2018ApJ...861L..13G} for PG~0136+251 and we ruled out its membership in both the Pleiades cluster and the moving group AB Doradus.  Although PG~0136+251, the Pleiades and AB Doradus share approximately the same common motion, we find that PG~0136+251 was actually further from the Pleiades and from AB Doradus in the past (independently of its line of sight velocity), so we excluded this WD from our sample as well.




\bibliography{main}

\bibliographystyle{aasjournal}



\end{document}